\documentclass[12pt]{article}
\usepackage[dvips]{graphics}

\begin{document}

\begin{center}

\LARGE{The mass and  spin\\ of the mesons, baryons and leptons}

\bigskip

\Large{E.L. Koschmieder}

\bigskip

\small{Center for Statistical Mechanics\\The University of Texas at Austin, Austin TX
78712.  USA\\
e-mail: koschmieder@mail.utexas.edu}

\end{center}
\bigskip

\noindent
\small
{The rest masses of the 
stable mesons and baryons and the rest masses of their antiparticles, as well as
 the rest masses of 
the $\mu^\pm$ and $\tau^\pm$ leptons can be explained, within 1\% accuracy,
 with the standing wave model, which uses only photons, neutrinos, charge and the 
weak nuclear force.  And we can explain the spin of the stable 
mesons and baryons and the spin of the $\mu^\pm$ and $\tau^\pm$ leptons
without any additional assumption. We can also determine the rest masses of the
\emph{e}, $\mu$ and $\tau$ neutrinos.}

\normalsize

\section*{Introduction}

 The so-called ``Standard Model" of the elementary particles 
has, until now, not come up with a precise theoretical determination of the
 masses of 
either the mesons and baryons or of the leptons, which means that neither 
the mass of the fundamental electron nor the mass of the fundamental 
proton have been explained. This is so although the quarks, the foundation 
of the standard model, have been introduced by Gell-Mann [1] forty 
years ago.
The masses of the quarks which are considered range from zero 
rest mass to values from 1.5 to 4.5\,MeV for e.g. the u-quark, according 
to the Review of Particle Physics [2], to values on the order of 100\,MeV. 
Suppose one 
has agreed on definite values of the masses of the various quarks then one 
stands before the same problem one has faced with the conventional 
elementary particles, namely one has to explain why the quarks have their 
particular masses and what they are made of. The other most frequently 
referred to theory dealing with the elementary particles, the ``String 
Theory" introduced about twenty years ago (Witten [3]), or its successor 
the superstring theory, have despite their mathematical elegance not led 
to experimentally verifiable results. There are many other attempts 
to explain the elementary particles or only one of the particles, too many 
to list them here. There have been, for example, in the last  
years, the articles of El Naschie on a general theory for high energy 
particles and the spectrum of the quarks [4-7]. Our model 
 and El Naschie's mechanical model version of his topological theory 
are not far apart.

   The need for the present investigation has been expressed by  
Feynman [8] as follows: ``There remains one especially unsatisfactory feature: the 
observed masses of the particles, m. There is no theory that adequately explains
 these numbers. We use the numbers in all our theories, but we do not understand
 them - what they are, or where they come from. I believe that from a fundamental
 point of view, this is a very interesting and serious problem". Today, twenty
 years later, we still stand in front of the same problem.   

\section {The spectrum of the masses of the particles}

As we have done before [9] we will focus attention on the so-called 
``stable" mesons and 
 baryons whose masses are reproduced with other data in Tables 1 and 2.
It is obvious that any attempt to explain the masses of the mesons and 
baryons should begin with the particles that are affected by the fewest 
parameters. These are certainly the particles without isospin (I = 0) and 
without spin (J = 0), but also with strangeness S = 0, and charm C = 0. 
Looking at the particles with I,J,S,C = 0 it is startling to find that 
their masses are quite close to integer multiples of the mass of the 
$\pi^0$\,meson. It is m($\eta) = (1.0140 \pm 0.0003)\,\cdot$\,4m($\pi^0$), 
and the mass of the resonance $\eta^\prime$ is m($\eta^\prime$) = (1.0137 
$\pm$ 0.00015)\,$\cdot$\,7m($\pi^0$). Three particles seem hardly to be 
sufficient to establish a rule. However, if we look a little further we 
find that m($\Lambda$) = 1.0332\,$\cdot$\,8m($\pi^0$) or m($\Lambda$) = 
1.0190\,$\cdot$\,2m($\eta$). We note that the $\Lambda$ particle has spin 1/2, 
not spin 0 as the $\pi^0$,\,$\eta$ mesons. Nevertheless, the mass of 
$\Lambda$ is close to 8m($\pi^0$). Furthermore we have m($\Sigma^0$) = 
0.9817\,$\cdot\,$9m($\pi^0$), m($\Xi^0) = 0.9742\,\cdot\,$10m($\pi^0$), 
m$(\Omega^-)$ = 
1.0325\,$\cdot\,$12m($\pi^0)$ = 1.0183\,$\cdot$\,3m($\eta$), ($\Omega^-$ 
is charged and has spin 3/2). Finally the masses of the charmed baryons are 
m($\Lambda_c^+$) = 0.9958\,$\cdot$\,17m($\pi^0$) = 
1.024\,$\cdot$\,2m($\Lambda$), m($\Sigma_c^0$) = 
1.0093\,$\cdot$\,18m($\pi^0$), 
m($\Xi_c^0$) = 1.0167\,$\cdot$\,18m($\pi^0$), and m($\Omega_c^0$) = 
1.0017\,$\cdot$\,20m($\pi^0$).
 
	\begin{table}\caption{The $\gamma$-branch of the particle 
spectrum}
	\begin{tabular}{lllllcl}\\
\hline\hline\\
 & m/m($\pi^0$) & multiples & decays & fraction & spin & 
mode\footnotemark[1]
 \\
 & & & & (\%) & &\\
[0.5ex]\hline
\\
$\pi^0$ & 1.0000 & 1.0000\,\,$\cdot$\,\,$\pi^0$ & $\gamma\gamma$ & 98.798 
& 0 & (1.)\\
 & & & $e^+e^-\gamma$ & \,\,\,1.198 & &\\
\\
$\eta$ & 4.0559 & 1.0140\,$\cdot$\,\,4$\pi^0$ & $\gamma\gamma$ & 39.43 & 0 
& 
(2.)\\
 & & & 3$\pi^0$ & 32.51 & &\\
 & & & $\pi^+\pi^-\pi^0$ & 22.6 & &\\
 & & & $\pi^+\pi^-\gamma$ & \,\,\,4.68 & &\\
\\
$\Lambda$ & 8.26577 & 1.0332\,$\cdot$\,\,8$\pi^0$ & p$\pi^-$ & 63.9 & 
$\frac{1}{2}$ 
& 2$\cdot$(2.)\\
 & & 1.0190\,$\cdot$\,\,2$\eta$ & n$\pi^0$ & 35.8 & &\\
\\
$\Sigma^0$ & 8.8352 & 0.9817\,$\cdot$\,\,9$\pi^0$ & $\Lambda \gamma$ & 100 
& 
$\frac{1}{2}$ & 2$\cdot (2.) + (1.)$\\
\\
$\Xi^0$ & 9.7417 & 0.9742\,$\cdot$\,10$\pi^0$ & $\Lambda\pi^0$ & 99.52 & 
$\frac{1}{2}$ & 
2$\cdot(2.) + 2(1.)$\\
\\
$\Omega^-$ & 12.390 & 1.0326\,$\cdot$\,12$\pi^0$ & $\Lambda$K$^-$ & 67.8 
& 
$\frac{3}{2}$ & 3$\cdot(2.)$\\
 & & 1.0183\,$\cdot$\,\,3$\eta$ & $\Xi^0\pi^-$ & 23.6 & &\\
 & & &  $\Xi^-\pi^0$ & \,\,\,8.6 & &\\
\\
$\Lambda_c^+$ & 16.928 & 0.9958\,$\cdot$\,17$\pi^0$ & many & & 
$\frac{1}{2}$ & 
2$\cdot(2.) + (3.)$\\
 & & 0.9630\,$\cdot$\,17$\pi^\pm$\\
\\
$\Sigma_c^0$ & 18.167 & 1.0093\,$\cdot$\,18$\pi^0$ & $\Lambda_c^+\pi^-$ & 
$\approx$100 & $\frac{1}{2}$ & $\Lambda_c^+ + \pi^-$\\
\\
$\Xi_c^0$ & 18.302 & 1.0167\,$\cdot$\,18$\pi^0$ & nine & (seen) & 
$\frac{1}{2}$   & 
2$\cdot(3.)$\\
\\
$\Omega_c^0$ & 20.033 & 1.0017\,$\cdot$\,20$\pi^0$ & six & (seen) & 
$\frac{1}{2}$ & 
2$\cdot(3.) + 2(1.)$\\
[0.2cm]\hline\hline
\vspace{0.1cm}
\end{tabular}
\footnotemark{\footnotesize The modes apply to neutral particles only. The $\cdot$
 sign marks coupled modes.}
	
	\end{table}

   Now we have enough material to 
formulate the $\emph{integer multiple}$ $\emph{rule}$, according to which 
the masses of the 
$\eta$,\,$\Lambda$,\,$\Sigma^0$,\,$\Xi^0$,
\,$\Omega^-$,\,$\Lambda_c^+$,\,$\Sigma_c^0$,
\,$\Xi_c^0$, and $\Omega_c^0$ 
particles are, in a first approximation, integer 
multiples of the mass of the $\pi^0$\,meson, although some of the particles 
have spin, and may also have charge as well as strangeness and charm. A 
consequence of the integer multiple rule must be that the ratio of the 
mass of any meson or baryon listed above divided by the mass of another 
meson or baryon listed above is equal to the ratio of two integer numbers. 
And indeed, for example m($\eta$)/m($\pi^0$) is practically two times 
(exactly 0.9950$\,\cdot$\,2) the ratio m($\Lambda$)/m($\eta$). There is 
also 
the ratio m($\Omega^-$)/m($\Lambda$) = 0.9993\,$\cdot$\,3/2. We have 
furthermore e.g. the ratios m($\Lambda$)/m($\eta$) = 1.019\,$\cdot$\,2, 
m($\Omega^-$)/m($\eta$) = 1.018\,$\cdot$\,3, m($\Lambda_c^+$)/m($\Lambda$) 
= 1.02399\,$\cdot$\,2, m($\Sigma_c^0$)/m($\Sigma^0$) = 1.0281\,$\cdot$\,2,  
m($\Omega_c^0$)/m($\Xi^0$) = 1.0282\,$\cdot$\,2, and 
m($\Omega_c^0$)/m($\eta$) = 0.9857\,$\cdot$\,5.

   We will call, for reasons to be explained later, the particles 
discussed above, which follow in a first approximation the integer 
multiple rule, the \emph{$\gamma$-branch} of the particle spectrum. The mass 
ratios of these particles are in Table 1. The deviation of the mass ratios 
from exact integer multiples of m($\pi^0$) is at most 3.3\%, the average 
of the factors before the integer multiples of m($\pi^0$) of the nine 
$\gamma$-branch particles in Table 1 is 1.0066 $\pm$ 0.0184. From a least 
square analysis follows that the masses of the ten particles on Table 1 
lie on a straight line given by the formula

\begin{equation} \mathrm{m}(N)/\mathrm{m}(\pi^0) = 1.0065\,N - 0.0043 
\qquad  
N\,\ge\,1, 
\end{equation}

\noindent     
where N is the integer number nearest to the actual ratio of the particle 
mass divided by m($\pi^0$). The correlation coefficient in equation (1) 
has the nearly perfect value r$^2$ = 0.999.
 
   The integer multiple rule applies to more than just the stable mesons 
and baryons. The integer multiple rule applies also to the $\gamma$-branch 
baryon resonances which have spin J = 1/2 and the meson resonances with 
I,J = 0, listed in [2] or in Tables 2,3 in [9]. The $\Omega^-$\,particle 
will not be considered because it has spin 3/2 but would not change the 
following equation significantly. If we combine the 
particles in Table 1 with the $\gamma$-branch meson and baryon 
resonances, that means if we consider all mesons and 
baryons of the $\gamma$-branch, ``stable" or unstable, with I $\leq$ 1, J 
$\leq$ 1/2 then we obtain from a least square analysis the formula 

\begin{equation}    \mathrm{m}(N)/\mathrm{m}(\pi^0) = 1.0056\,N + 
0.0610\qquad    
N \ge 1,  
\end{equation}

\noindent
with the correlation coefficient 0.9986. The line through the points is 
shown in Fig.\,1 which tells that 22 particles of the $\gamma$-branch of 
different spin and isospin, strangeness and charm; eight I,J = 0,0 mesons, 
thirteen J = 1/2 
baryons and the $\pi^0$\,meson with I,J = 1,0, lie on a straight line with 
slope 1.0056. In other words they approximate the integer multiple rule 
very well.

\begin{figure}[h] 
	\vspace{0.5cm}
	\includegraphics{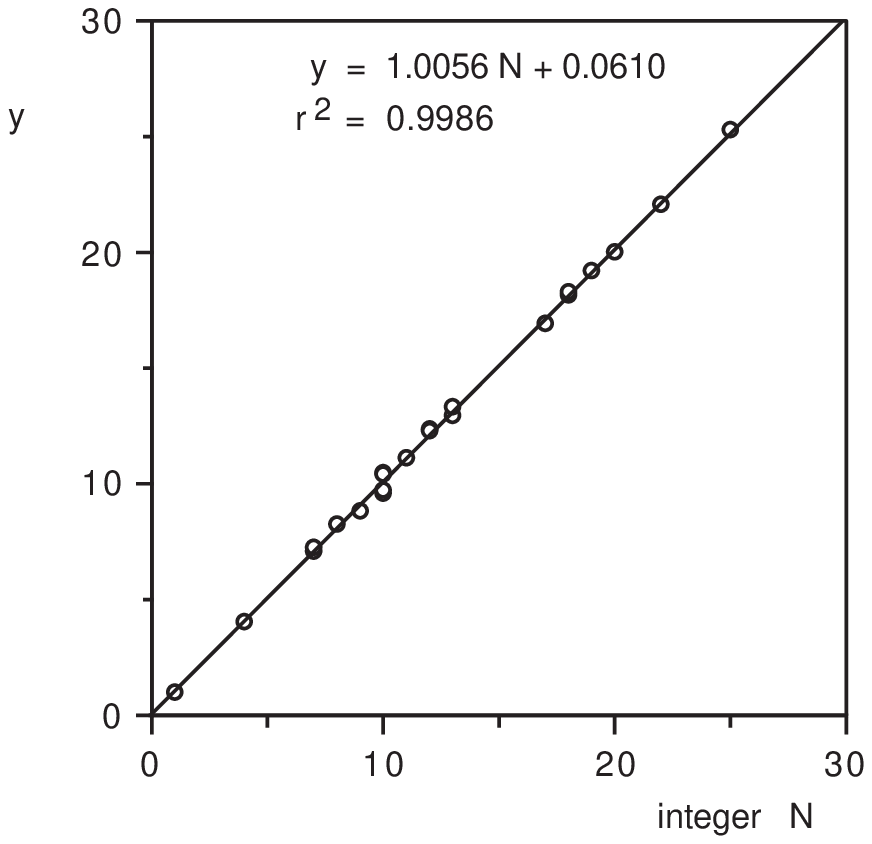}
	\vspace{-0.3cm}
	\begin{quote}
Fig.\,1: The mass of the mesons and baryons of the $\gamma$-branch with 
I\,$\leq$ 1,\,J\,$\leq\frac{1}{2}$ in units of m($\pi^0$) as a function of 
the integer  N. y = m/m($\pi^0$).
	\end{quote}
\end{figure}

\noindent

   Searching for what else the 
$\pi^0$,\,$\eta$,\,$\Lambda$,\,$\Sigma^0$,\,$\Xi^0$,\,$\Omega^-$ particles 
have 
in common, we find that the principal decays (decays with a 
fraction\,$>$\,1\%) 
of these particles, as listed in Table 1, involve primarily $\gamma$-rays, 
the characteristic case is $\pi^0 \rightarrow \gamma\gamma$  (98.8\%). We 
will later on discuss a possible explanation for the 1.198\% of the decays 
of $\pi^0$ which do not follow the $\gamma\gamma$ route. After the 
$\gamma$-rays the next most 
frequent decay product of the heavier particles of the $\gamma$-branch are 
$\pi^0$\,mesons which again decay into $\gamma\gamma$. To describe the 
decays in another way, the principal decays of the particles listed above 
take place $\emph{always without the emission of neutrinos}$\,; see Table 1. 
There the decays and the fractions of the principal decay modes are given, 
taken from the Review of Particle Physics. We cannot consider decays 
with fractions $< 1\%$. We will refer to the particles whose masses are 
approximately integer multiples of the mass of the $\pi^0$\,meson, and 
which decay without the emission of neutrinos, as the 
$\gamma$-$\emph{branch}$ of the particle spectrum.

   To summarize the facts concerning the $\gamma$-branch. Within 
0.66\% on the average 
the masses of the particles of the $\gamma$-branch are integer multiples 
(namely 4,\,8,\,9,\,10,\,12, and even 17,\,18,\,20) of the mass of the 
$\pi^0$\,meson. 
It is improbable that nine particles have masses so close to integer 
multiples of m($\pi^0$) if there is no correlation between them and the 
$\pi^0$\,meson. It has, on the other hand, been argued that the integer 
multiple rule is a numerical coincidence. But the probability that the 
mass ratios of the $\gamma$-branch fall by coincidence on integer numbers 
between 1 and 20 instead on all possible numbers between 1 and 20 with two 
decimals after the period is smaller than 10$^{-20}$, i.e nonexistent. The 
integer multiple rule is not affected by more than 3\% by the spin, the 
isospin, the strangeness, and by charm. The integer multiple rule seems 
even to apply to the $\Omega^-$ and $\Lambda_c^+$ particles, although they 
are charged. In order for the integer multiple rule to be valid the 
deviation of the ratio m/m($\pi^0$) from an integer number must be smaller 
than 1/2N, where N is the integer number closest to the actual ratio 
m/m($\pi^0$). That means that the permissible deviation decreases rapidly 
with increased N. All particles of the $\gamma$-branch have deviations 
smaller than 
1/2N.
 
   The remainder of the stable mesons and baryons are the  
$\pi^\pm$, K$^{\pm,0}$,\,p,\,\,n,\quad D$^{\pm,0}$, and D$_s^\pm$ 
particles which make up 
the \emph{$\nu$-branch} of the particle spectrum. The ratios of their masses 
are given in Table 2.

	\begin{table}\caption{The $\nu$-branch of the particle 
spectrum}
	\begin{tabular}{llllrcl}\\

\hline\hline\\
 & m/m($\pi^\pm$) & multiples & decays\footnotemark[2]
 & fraction & spin & mode 
 \\
 & & & & (\%) & &\\
[0.5ex]\hline
\\
$\pi^\pm$ & 1.0000 & 1.0000\,\,$\cdot$\,\,$\pi^\pm$ & $\mu^+\nu_\mu$ & 
\,\,99.9877 & 0 & (1.)\\
\\
K$^{\pm,0}$ & 3.53713 & 0.8843\,$\cdot$\,\,4$\pi^\pm$ & $\mu^+\nu_\mu$ & 
\,\,63.43 & 0 & 
(2.) + $\pi^0$\\
 & & & $\pi^\pm\pi^0$ & \,\,21.13 & &\\
 & & & $\pi^+\pi^-\pi^+$ & \,\,\,\,\,5.58 & &\\
 & & & $\pi^0 e^+ \nu_e$ (K$_{e3}^+$)& \,\,\,\,\,4.87 & &\\
 & & & $\pi^0\mu^+\nu_\mu$ (K$_{\mu3}^+$) & \,\,\,\,\,3.27 & &\\
\\
n & 6.73186 & 0.8415\,$\cdot$\,\,8$\pi^\pm$ & $p\,e^-\overline{\nu}_e$ & 
100.\,\,\,\,\,\,\,\,  & $\frac{1}{2}$ & 2$\cdot$(2.) + $2\pi^\pm$\\
 & & 0.9516\,$\cdot$\,\,2K$^\pm$ \\
\\
D$^{\pm,0}$ & 13.393 & 0.8370\,$\cdot$\,16$\pi^\pm$ & $e^+$ anything & 
\,\,17.2 & 
0 & 2(2$\cdot(2.) + 2\pi^\pm$)\\
 & & 0.9466\,$\cdot$\,\,4K$^\pm$ & K$^-$ anything & \,\,24.2 \\
 & & 0.9954\,$\cdot$\,(p + $\bar{\mathrm{n}}$) & $\overline{\mathrm{K}}^0$ anything\\	
 & &                       & \,\,+\,K$^0$ anything & \,\,59\\
 & &                       & $\eta$ anything & $<$\,13\\
&&                          &K$^+$ anything &5.8\\
\\
D$^\pm_s$ & 14.104 & 0.8296\,$\cdot$\,17$\pi^\pm$ & K$^-$ anything & 
\,\,13 & 0 & 
body centered \\
 & & 0.9968\,$\cdot$\,\,4K$^\pm$ & $\overline{\mathrm{K}}^0$ 
 anything & & & \,\, cubic\\
 & &                            & \,\,+\,K$^0$ anything & \,\,39\\
 & &                            & K$^+$ anything & \,\,20\\
 & &                            & $e^+$ anything & 8\\
[0.2cm]\hline\hline

\vspace{0.1cm}
	\end{tabular}

	\footnotemark{\footnotesize The particles with negative 
	charges have conjugate charges of the listed  decays. Only the 
	decays of K$^\pm$ and D$^\pm$ are listed. The oscillation modes carry 
one                electric charge.}

	\end{table}

These particles are in general charged, exempting the K$^0$ and D$^0$ 
mesons and the neutron n, in contrast to the particles of the 
$\gamma$-branch, which are in general neutral. It does not make a 
significant difference whether one considers the mass of a particular 
charged or neutral particle. After the $\pi$\,mesons, the largest mass 
difference between charged and neutral particles is that of the K\,mesons 
(0.81\%), and thereafter all mass differences between charged and neutral 
particles are $<0.5\%$. The integer multiple rule does not immediately 
apply to the masses of the $\nu$-branch particles if m($\pi^\pm$) (or 
m($\pi^0$)) is used as reference, because m(K$^\pm$) = 
0.8843\,$\cdot$\,4m($\pi^\pm$). 0.8843\,$\cdot$\,4 = 3.537 is far from 
integer. Since 
the masses of the $\pi^0$\,meson and the $\pi^\pm$\,mesons differ by only 
3.4\% it has been argued that the $\pi^\pm$\,mesons are, but for the 
isospin, the same particles as the $\pi^0$\,meson, and that therefore the 
$\pi^\pm$ cannot start another particle branch. However, this argument is 
not supported by the completely different decays of the $\pi^0$\,mesons 
and 
the $\pi^\pm$\,mesons. The $\pi^0$\,meson decays almost exclusively into 
$\gamma\gamma$ (98.8\%),  whereas the $\pi^\pm$\,mesons decay practically 
exclusively into $\mu$\,mesons and neutrinos, as in $\pi^+$ $\rightarrow$  
$\mu^+$  + $\nu_\mu$  (99.9877\%). Furthermore, the lifetimes of the 
$\pi^0$  and the $\pi^\pm$ mesons differ by nine orders of magnitude, 
being $\tau$($\pi^0$) = 8.4\,$\cdot$\,10$^{-17}$ sec versus 
$\tau$($\pi^\pm$) 
= 2.6\,$\cdot$\,10$^{-8}$ sec.

   If we make the $\pi^\pm$\,mesons the reference particles of the 
$\nu$-branch, then we must multiply the mass ratios m/m($\pi^\pm$) of the 
above listed particles with an average factor 0.848 $\pm$ 0.025, as 
follows from the mass ratios on Table 2. 
The integer multiple rule may, however, apply 
directly if one makes m(K$^\pm$) the reference for masses larger than 
m(K$^\pm$). The mass of the neutron is 0.9516\,$\cdot$\,2m(K$^\pm$),
 which is only a fair approximation to an integer multiple. There are, on the other 
hand, outright integer multiples in m(D$^\pm$) = 0.9954\,$\cdot$\,(m(p) + 
m(n)), and in m(D$_s^\pm$) = 0.9968\,$\cdot$\,4m(K$^\pm$). A least square 
analysis of the masses of the $\nu$-branch in Table 2 yields the formula

\begin{equation} \mathrm{m}(N)/0.853\mathrm{m}(\pi^\pm) = 1.000\,N + 
0.00575\qquad  N \ge 1 ,
\end{equation}
\noindent
with r$^2$ = 0.998. This means that the particles of the $\nu$-branch are 
integer multiples of m($\pi^\pm$) times the factor 0.853. One must, 
however, consider that the $\pi^\pm$\,mesons are not necessarily the 
perfect reference for all $\nu$-branch particles, because $\pi^\pm$ has I 
= 1, whereas for example K$^\pm$ has I = 1/2 and S = $\pm$1 and the 
neutron has also I = 1/2. Actually the 
factor 0.853 in Eq.(3) is only an average. The mass ratios indicate that this factor 
 decreases slowly with increased m(N).
 The existence of the factor and its decrease will be explained later.

   Contrary to the particles of the $\gamma$-branch, the $\nu$-branch 
particles decay preferentially with the emission of neutrinos, the 
foremost example is $\pi^\pm \rightarrow \mu^\pm$ + 
$\nu_\mu(\bar{\nu}_\mu)$ with a 
fraction of 99.9877\%. Neutrinos characterize the weak interaction. We will 
refer to the particles in Table 2 as the $\emph{neutrino branch}$ 
($\nu$-branch) of the particle spectrum. We emphasize that a weak decay of 
the particles of the $\nu$-branch is by no means guaranteed. Although the 
neutron decays via n $\rightarrow$ p + e$^-$ + $\bar{\nu}_e$ in 887 sec 
(100\%), the proton is stable. There are, on the other hand, decays as 
e.g. K$^+ 
\rightarrow \pi^+\pi^-\pi^+$ (5.59\%), but the subsequent decays of the 
$\pi^\pm$\,mesons lead to neutrinos and e$^\pm$. The decays of the 
particles in the 
$\nu$-branch follow a mixed rule, either weak or electromagnetic.

   To summarize the facts concerning the $\nu$-branch of the mesons and 
bary-ons. The masses of these particles seem to follow the integer 
multiple 
rule if one uses the $\pi^\pm$\,meson as reference, however the mass 
ratios share a common factor 0.85 $\pm$ 0.025.

   To summarize what we have learned about the integer multiple rule: In 
spite of differences in charge, spin, strangeness, and 
charm the masses of the mesons and baryons of the $\gamma$-branch are 
 integer multiples 
of the mass of the $\pi^0$\,meson within at most 3.3\% and on the average 
within 0.66\%. Correspondingly, the masses of the 
particles of the $\nu$-branch are, after multiplication with a factor 0.85 
$\pm$ 0.025, integer multiples of the mass of the $\pi^\pm$\,mesons. The 
validity of the integer multiple rule can easily be verified with a 
calculator from the data in the Review of
Particle Physics. The integer multiple rule has been anticipated 
much earlier by Nambu [10], who wrote in 1952 that ``some regularity 
[in the masses of the particles] might be found if the masses were 
measured in a unit of the order of the $\pi$-meson mass". A similar 
suggestion has been made by Fr\"ohlich [11]. The integer multiple rule 
suggests that the particles are the result of superpositions of modes and 
higher modes of a wave equation.

\section {Standing waves in a cubic lattice and the particles of the 
$\gamma$-branch} 

 We will now study, as we 
have done in [12], whether the so-called ``stable" particles of the 
$\gamma$-branch cannot be described by the frequency spectrum of standing 
waves in a cubic lattice, which can accommodate automatically the Fourier 
frequency spectrum of an extreme short-time collision by which the 
particles are created. 
The investigation of the consequences of lattices for particle theory 
was initiated by Wilson [13] who studied a cubic fermion lattice. This study has 
developed over time into lattice QCD. 

   It will be necessary for the following to outline the most elementary aspects
 of the theory of lattice oscillations. The classic paper describing lattice oscillations
 is from 
Born and v.\,Karman [14], henceforth referred to as B\&K. They looked at first 
at the oscillations of a one-dimensional chain of points with mass m, 
separated by a constant distance $\emph{a}$. This is the 
$\emph{monatomic}$ case, all lattice points have the same mass. B\&K 
assume 
that the forces exerted on each point of the chain originate only from the 
two neighboring points. These forces are opposed to and proportional to 
the displacements, as with elastic springs (Hooke's law). The equation of 
motion is in this case

\begin{equation} \mathrm{m}\ddot{u}_{n} = \alpha(u_{n+1} - u_n) - \alpha(u_n - 
u_{n-1})\,.
\end{equation}
\noindent
The $u_n$ are the displacements of the mass points from their equilibrium 
position 
which are apart by the distance \emph{a}. The dots signify, as usual, 
differentiation with respect 
to time, $\alpha$ is a constant characterizing the force between the 
lattice points, and n is an integer number. For \emph{a} $\rightarrow$ 0 
Eq.(4) becomes the wave equation c$^2\partial^2$u/$\partial$x$^2$ = 
$\partial^2$u/$\partial$t$^2$ (B\&K).
 
   In order to solve (4) B\&K set 

\begin{equation} u_n = Ae^{i(\omega\,t\, +\, n\phi)}\,,
\end{equation}

\noindent
which is obviously a temporally and spatially periodic solution. n is an 
integer, with n  $<$ N, where N is the number of points in the chain. 
$\phi$ = 0 is the monochromatic case. We also consider higher modes,
 by replacing n$\phi$ in Eq.(5) with
$\ell$n$\phi$, with integer $\ell$\,$>$\,1. The wavelengths are then shorter by
 1/$\ell$. At n$\phi$ = $\pi$/2 there are 
nodes, where for all times $\emph{t}$ the displacements are zero, as with 
standing 
waves. If a displacement is repeated after n points we have n$\emph{a}$ = 
$\lambda$, where $\lambda$ is the wavelength, $\emph{a}$ the lattice 
constant, and it must be n$\phi$ = 2$\pi$ according to (5). It follows that

\begin{equation} \lambda = 2\pi\emph{a}/\phi\,.
\end{equation}
 
Inserting (5) into (4) one obtains a continuous frequency spectrum given 
by Eq.(5) of B\&K

\begin{equation}  \omega = \pm\,2\sqrt{\alpha/\mathrm{m}}\,\mathrm{sin}(\phi/2)\,.
\end{equation}
\noindent
B\&K point out that there is not only a continuum of frequencies, but also 
a \emph{maximal frequency} which is reached when $\phi$ = $\pi$, or at the 
minimum of the possible wavelengths $\lambda$ = 2$\emph{a}$. The 
boundary conditions are periodic, that means that $u_n$ = $u_{n + N}$, 
where N is the number of points in the chain. Born referred to the 
periodic boundary condition as a ``mathematical convenience". The number 
of normal modes must be equal to the number of particles in the lattice.
 
    Born's model of the crystals has been verified in great detail by X-ray
 scattering and even in much more complicated cases by neutron scattering. 
 The theory of lattice oscillations has been pursued in particular by 
Blackman [15], a summary of his and other studies is in [16]. 
Comprehensive reviews of the results of linear studies of lattice dynamics 
have been written by Born and Huang [17], by Maradudin et al. [18], and by 
Ghatak and Kothari [19].

\section {The masses of the $\gamma$-branch particles}

   We will now assume, as seems to be quite natural, that the particles
 \emph{consist of the same particles into which they 
decay}, directly or ultimately. We know this from atoms, which consist of
nuclei and electrons, and from nuclei, which consist of protons and neutrons. 
Quarks have never appeared among the decay products of elementary particles.
For the $\gamma$-branch 
particles our assumption means that they consist of photons. Photons 
and $\pi^0$\,mesons are the principal decay products of the 
$\gamma$-branch particles, the 
characteristic example is $\pi^0$ $\rightarrow$ $\gamma\gamma$  (98.8\%). 
Table 1 shows that there are decays of the $\gamma$-branch particles which 
lead to particles of the $\nu$-branch, in particular to pairs of $\pi^+$ 
and $\pi^-$ mesons.  It appears 
that this has to do with pair production in the $\gamma$-branch particles. 
Pair production is evident in the decay $\pi^0 \rightarrow e^+ + e^- + 
\gamma$ (1.198\%) or in the $\pi^0$\,meson's third most frequent decay
 $\pi^0 \rightarrow e^+e^-e^+e^-$ (3.14$\cdot10^{-3}$\%). Pair production
 requires the presence of electromagnetic waves of high 
energy. Anyway, the explanation of the $\gamma$-branch particles must begin 
with the explanation of the most simple example of its kind, the 
$\pi^0$\,meson, which by all means seems to consist of photons. 
The composition of the particles of the $\gamma$-branch suggested here 
offers a direct route from the formation of a $\gamma$-branch particle, 
through its lifetime, to its decay products. Particles that are made of 
photons are necessarily neutral, as the majority of the particles of the 
$\gamma$-branch are.
 
   We also base our assumption that the particles of the $\gamma$-branch 
are made of photons on the circumstances of the formation of the 
$\gamma$-branch particles. The most simple and straightforward creation of 
a $\gamma$-branch particle are the reactions $\gamma$ + p $\rightarrow$ 
$\pi^0$ + p, or in the case that the spins of $\gamma$ and p are parallel 
$\gamma$ + p $\rightarrow$ 
$\pi^0$ + p + $\gamma^\prime$. A photon impinges on a proton and creates a 
$\pi^0$\,meson. The considerations which follow apply as well for other
 photoproductions such as $\gamma$ + p $\rightarrow \eta$ + p or $\gamma$ + d 
 $\rightarrow \pi^0$ + d, but also for the electroproductions e$^-$ + p $\rightarrow 
\pi^0$ + e$^-$ + p or e$^-$ + d $\rightarrow$   $\pi^0$  + e$^-$ + d, see Rekalo
 et al. [20]. 

   In $\gamma + p \rightarrow \pi^0 + p$   the pulse of the 
incoming electromagnetic wave is in 10$^{-23}$\,sec  converted 
into a continuum of electromagnetic waves with frequencies ranging from 
10$^{23}$ sec$^{-1}$ to $\nu$ $\rightarrow$ $\infty$ according to Fourier 
analysis. There must be a 
cutoff frequency, otherwise the energy in the sum of the frequencies would 
exceed the energy of the incoming electromagnetic wave. The wave packet so 
created decays, according to experience, after 8.4\,$\cdot$\,10$^{-17}$\,sec 
into two electromagnetic waves or $\gamma$-rays. It seems to be very 
unlikely that Fourier analysis does not hold for the case of an 
electromagnetic wave impinging on a proton. The question then arises of 
what happens to the electromagnetic waves in the timespan of 10$^{-16}$ 
seconds between the creation of the wave packet and its decay into two 
$\gamma$-rays? We will investigate whether the electromagnetic waves 
cannot continue to exist for the 10$^{-16}$ seconds until the wave packet 
decays.

   If the wave packet created by the collision of a $\gamma$-ray with a proton
consists of electromagnetic waves, then the waves cannot be progressive
because the wave packet must have a \emph{rest mass}. However \emph{standing electromagnetic waves} can have a rest mass. Standing electromagnetic waves 
are equivalent to a lattice and the lattice oscillations can absorb the continuum of
 frequencies of the Fourier spectrum of the collision. So we assume that the
 very many photons in the wave packet are held together in a cubic lattice. It is not unprecedented that photons have been 
considered to be building blocks of the elementary particles. Schwinger 
[21] has once studied an exact one-dimensional quantum electrodynamical 
model in which the photon acquired a mass $\sim$ $e^2$. On the other hand,
 it has been suggested by Sidharth [22] that the  $\pi^0$ meson 
consists of an electron and a positron which circle their center of mass.

   We will now investigate the standing waves of a cubic photon lattice. 
We assume that the lattice is held together by a weak force acting from 
one lattice point to the next. We assume that the range of this force is 
10$^{-16}$\,cm, because the range of the weak nuclear force is on the 
order 
of 10$^{-16}$\,cm, according to Perkins [23]. The range of the weak force is of 
the same magnitude as the uncertainty $\Delta$x = $\emph{a}$/$\pi$ of 
the location of a 
wavepacket whose energy E is $\approx$ $\pi$h$\nu_0/2$ = 
$\pi$/2$\cdot$hc/2$\pi\emph{a}$, where the energy E is the average energy 
of the photons in a lattice with the lattice constant \emph{a}, as we 
will see later. For the sake of simplicity we set the 
sidelength of the lattice at 10$^{-13}$\,cm, the exact size of the nucleon 
is given in [24] and will be used later. With \emph{a} = $10^{-16}$\,cm 
there are then 10$^9$ lattice 
points. As we will see the ratios of the masses of the photon lattices are 
independent of the 
sidelength of the lattice. Because it is the most simple case, we assume 
that a central force acts between the lattice points. We cannot consider 
spin, isospin, strangeness or charm of the particles. The frequency 
equation for the oscillations of an isotropic monatomic cubic lattice with 
central forces is, in the one-dimensional case, given by Eq.(7). The 
direction of the oscillation is determined by the direction of the 
incoming wave.

   According to Eq.(13) of B\&K the force constant $\alpha$ is

\begin{equation}   \alpha = \emph{a}(c_{11} - c_{12} - c_{44})\,,
\end{equation}

\noindent
where c$_{11}$, c$_{12}$ and c$_{44}$ are the elastic constants in 
continuum mechanics which applies in the limit $\emph{a}$ $\rightarrow$  
0. If we consider central forces then c$_{12}$ = c$_{44}$ which is the 
classical Cauchy relation. Isotropy requires that c$_{44}$ = 
(c$_{11}\,-\,$  c$_{12}$)/2.  
The waves are longitudinal. Transverse waves in a cubic lattice 
with concentric forces are not possible according to [19]. All frequencies 
that solve Eq.(7) come with either a plus or a minus sign which is, as we 
will see, important. The reference frequency in Eq.(7) is

\begin{equation}  \nu_0 = \sqrt{\alpha/4\pi^2\mathrm{m}}\,,
\end{equation}

\noindent
or as we will see, using Eq.(11),  $\nu_0$  =  c$_\star/2\pi\emph{a}$.

   The \emph{limitation of the group velocity} in the photon lattice has 
now to be considered. The group velocity is given by

\begin{equation}  c_g = \frac{d\omega}{dk} = 
\emph{a}\sqrt{\frac{\alpha}{\mathrm{m}}}\cdot\frac{df(\phi)}{d\phi}\,.
\end{equation}
\noindent
The group velocity in the photon lattice has to be equal to the 
\emph{velocity of light} c$_\star$ throughout the entire frequency spectrum, because 
photons move with the velocity of light. In order to learn how this 
requirement affects the frequency distribution we have to know the value 
of $\sqrt{\alpha/\mathrm{m}}$ in a photon lattice. But we do not have information 
about what either $\alpha$ or m might be in this case. We assume in the 
following that $\emph{a}\sqrt{\alpha/\mathrm{m}}$ = c$_\star$, which means, since 
$\emph{a} = 10^{-16}$\,cm, that $\sqrt{\alpha/\mathrm{m}}$ = 3\,$\cdot$\,10$^{26}$ 
sec$^{-1}$, or that the corresponding period is $\tau$ = 
1/3\,$\cdot$\,10$^{-26}$ sec, which is the time it takes for a wave to 
travel with the velocity of light over one lattice distance. With 

\begin{equation} c_\star = \emph{a}\sqrt{\alpha/\mathrm{m}}\,
\end{equation}

\noindent
the equation for the group velocity is

\begin{equation} c_g = c_\star\cdot df/d\phi\,.
\end{equation} 

  For a photon lattice that means, since c$_g$ must then always be equal 
to c$_\star$, that df/d$\phi$ = 1. This requirement determines the form of 
the frequency distribution regardless of the 
order of the mode of oscillation or it means that instead of the sine 
function in Eq.(7) the frequency is given by

\begin{equation}\nu = \pm\,\nu_0[\,\phi + \phi_0\,]\,. \end{equation} 

 For the time being we will disregard $\phi_0$ in Eq.(13). The frequencies
 of the corrected 
spectrum must increase from $\nu$ = 0 at the origin $\phi$ = 0 with slope 
1 (in units of $\nu_0$) until the maximum is reached at $\phi = \pi$. 
The energy  contained in the oscillations (Eq.14) must be proportional to
 the sum of all frequencies.
The \emph{second mode}  of the lattice oscillations contains 4 times as much 
energy as  the basic mode, because the frequencies are twice the 
frequencies of the basic mode, and there are twice as many oscillations. 
 Adding, by superposition, to the second mode different 
numbers of basic modes or of second modes will give exact integer 
multiples 
of the energy of the basic mode. Now we understand the integer multiple 
rule of the particles of the $\gamma$-branch. There is, in the framework 
of this theory, on account of Eq.(13), no alternative but $\emph{integer 
multiples}$ of the basic mode for the energy contained in the frequencies 
of the different modes or for superpositions of different modes. In other 
words, the masses of the different particles are integer multiples of the 
mass of the $\pi^0$\,meson, assuming that there is no spin, isospin, 
strangeness or charm.

   We remember that the measured masses in Table 1, 
which incorporate different spins, isospins, strangeness, and charm spell 
out the integer multiple rule within on the average 0.65\% accuracy. It is 
worth noting 
that $\emph{there is no free parameter}$ if one takes the ratio of the 
energies contained in the frequency distributions of the different modes, 
because the factor $\sqrt{\alpha/\mathrm{m}}$ in Eq.(7) cancels. This means, in 
particular, that the ratios of the frequency distributions, or the mass 
ratios, are independent of the mass of the photons at the lattice points, 
as well as of the magnitude of the force between the lattice points.

   It is obvious that the integer multiples of the frequencies are only a 
first approximation of the theory of lattice oscillations and of the mass 
ratios of the particles. The equation of motion in the lattice (4) does 
not apply in the eight corners of the cube, nor does it 
apply to the twelve edges nor, in particular, to the six sides of the 
cube. A cube with 10$^9$ lattice points is not correctly described by the 
periodic boundary condition we have used, but is what is referred to as a 
microcrystal. A phenomenological theory of the frequency distributions in 
microcrystals, considering in particular surface waves, can be found in 
Chapter 6 of Ghatak and Kothari [19]. Surface waves may account for the 
small deviations of the mass ratios of the mesons and baryons from the 
integer multiple rule of the oscillations in a cube. However, it seems to 
be futile to 
pursue a more accurate determination of the oscillation frequencies as 
long as one does not know what the structure of the electron is, whose 
mass is 0.378\% of the mass of the $\pi^0$\,meson and hence is a 
substantial part of the deviation of the mass ratios from the integer 
multiple rule.
 
   Let us summarize our findings concerning the particles of the 
$\gamma$-branch. The $\pi^0$\,meson is the basic mode of the
 photon lattice 
oscillations. The $\eta$\,meson corresponds to the second oscillation mode,
 as is suggested by m($\eta$) $\approx$ 4m($\pi^0$). The 
$\Lambda$ particle corresponds to the superposition of two second modes,
 as is suggested by m($\Lambda) \approx$ 2m($\eta$). This 
superposition apparently results in the creation of spin 1/2. The two 
modes would then have to be coupled. The $\Sigma^0$ and $\Xi^0$ baryons 
are superpositions of one or two basic modes on the $\Lambda$  particle. 
The $\Omega^-$ particle corresponds to the superposition of three coupled 
second modes  as is suggested by m($\Omega^-$) $\approx$ 
3m($\eta$). This procedure apparently causes spin 3/2. The charmed 
$\Lambda_c^+$ particle seems to be the first particle incorporating a 
third oscillation mode. $\Sigma_c^0$ is apparently the superposition of a 
negatively charged basic mode 
on $\Lambda_c^+$, as is suggested by the decay of $\Sigma_c^0$. The 
easiest explanation of $\Xi_c^0$ is that it is the superposition of two 
coupled third modes. The superposition of two modes of the same type is, 
as in the case of $\Lambda$, accompanied by spin 1/2. The $\Omega_c^0$ 
baryon is apparently the superposition of two basic modes on the $\Xi_c^0$ 
particle. All neutral particles of the $\gamma$-branch are thus accounted 
for. The modes of the particles are listed in Table 1.

   We have also found the $\gamma$-branch $\emph{antiparticles}$, which 
follow from the negative frequencies which solve Eq.(7). 
Antiparticles have always been associated with negative energies. 
Following Dirac's argument for electrons and positrons, we associate the 
masses with the negative frequency distributions with 
antiparticles. We emphasize that the existence of antiparticles is an 
automatic consequence of our theory.

   All particles of the $\gamma$-branch are unstable with lifetimes on the 
order of 10$^{-10}$ sec or shorter. Born [25] has shown that the 
oscillations in cubic lattices held together by central forces are 
unstable. It seems, however, to be possible that the particles can be 
unstable for reasons other than the instability of the lattice. For 
example, pair production seems to make it possible to understand the decay 
of the 
$\pi^0$\,meson $\pi^0 \rightarrow e^- + e^+ + \gamma$  (1.198\%).  Since 
in our model the $\pi^0$\,meson consists of a multitude of electromagnetic 
waves it seems 
that pair production takes place within the $\pi^0$\,meson, and even more so 
in the higher modes of the $\gamma$-branch where the electrons and 
positrons 
created by pair production tend to settle on mesons, as e.g. in $\eta 
\rightarrow \pi^+ + 
\pi^- + \pi^0$ (22.6\%) or in the decay $\eta \rightarrow \pi^+ + \pi^- + 
\gamma$ (4.68\%), where the origin of the pair of charges is more apparent.
Pair production is also evident in the decays $\eta \rightarrow 
e^+e^-\gamma$ (0.6\%)  or $\eta \rightarrow e^+e^-e^+e^-$ 
(6.9$\cdot10^{-3}$\%).

   Finally we must explain the reason for which the photon lattice or the 
$\gamma$-branch particles are limited in size to a particular value of 
about $10^{-13}$ cm, as 
the experiments tell. Conventional lattice theory using the periodic 
boundary condition does not limit the size of a crystal, and in fact very 
large crystals exist. If, however, the lattice consists of standing 
electromagnetic waves the size of the lattice is limited by the radiation 
pressure. The lattice will necessarily break up at the latest when the 
outward directed radiation pressure is equal to the inward directed 
elastic force which holds the lattice together. For details we refer to [26].

\section {The mass of the $\pi^0$\,meson}

   So far we have studied the ratios of the masses of the particles. We 
will now determine the mass of the $\pi^0$\,meson in order to validate 
that 
the mass ratios link with the actual masses of the particles. The 
energy of the $\pi^0$\,meson is \vspace{0.5cm}

\centerline{E(m($\pi^0$)) = 134.9766\,MeV = 2.1626\,$\cdot\,10^{-4}$\,erg.}
\vspace{0.5cm}
\noindent
For the sum of the energies  of the frequencies of all standing one-dimensional
waves in $\pi^0$  we use the equation

\begin{equation}\mathrm{E}_\nu = 
\frac{\mathrm{Nh}\nu_0}{2\pi}\,\,\int\limits_{-\pi}^{\pi}\,f(\phi)d\phi\,.
\end{equation}

  This equation originates from B\&K. N is the number of all lattice 
points. The total energy of the frequencies in a cubic lattice is equal 
to the number N of the oscillations times the average of the energy of the 
individual frequencies. In order to arrive at an exact value of Eq.(14) 
we have to use the correct value of the radius of the proton, which is 
r$_p$ = (0.880 $\pm$ 0.015)\,$\cdot$\,10$^{-13}$\,cm according to [24] or
r$_p$ = (0.883 $\pm$  0.014)\,$\cdot$\,10$^{-13}$\,cm according to [27].  
With $\emph{a}$ = 10$^{-16}$\,cm it follows that the number of all lattice 
points in the cubic lattice is

\centerline{N = 2.854\,$\cdot$\,10$^9$.}
\noindent
The radius of the $\pi^\pm$\,mesons has also been measured [28] and after 
further analysis [29] was found to be 0.83\,$\cdot$\,10$^{-13}$\,cm, which 
means that within the uncertainty of the radii we have $r_p$ = $r_\pi$. 
And according to [30] the charge radius of $\Sigma^-$ is (0.78 $\pm$ 
0.10)\,$\cdot$\,$10^{-13}$\,cm.

  If the oscillations are parallel to an axis, the 
limitation of the group velocity is taken into account, that means if 
Eq.(13) applies and the absolute 
values of the frequencies are taken, then the value of the 
 integral in Eq.(14) is $\pi^2$. With N = 
2.854\,$\cdot$\,10$^9$ and $\nu_0$ = c$_\star/2\pi\emph{a}$ it follows from 
Eq.(14) that the sum of the energy of the frequencies corrected for the 
group velocity limitation of the basic mode  is 
E$_{corr}$ = 1.418$\cdot$10$^9$\,erg. That means  
that the energy is 6.56$\cdot10^{12}$ times larger than E(m($\pi^0$)). 
This discrepancy is inevitable, because 
the basic frequency of the Fourier spectrum after a collision on the order of 
10$^{-23}$ sec duration is $\nu$ = 10$^{23}$ sec$^{-1}$, which means, when 
E = h$\nu$, that one basic frequency alone contains an energy of about 
9\,m($\pi^0$)c$_\star^2$.
 
   To eliminate this discrepancy we use, instead of the simple 
form E = h$\nu$, the complete quantum mechanical energy of a linear 
oscillator as given by Planck

\begin{equation} E = \frac{h\nu}{e^{h\nu/kT} -\,1}\,\,.
\end{equation}
\noindent
This equation was already used by B\&K for the determination of the 
specific heat of cubic crystals or solids. Equation (15) calls into 
question the value of the temperature T in the interior of a particle. We 
determine T empirically with the formula for the internal energy of solids

\begin{equation} u = \frac{R\Theta}{e^{\Theta/T} -1}\,\,,
\end{equation}
\noindent
which is from Sommerfeld [31]. In this equation R = 
2.854\,$\cdot\,10^9$\,k, where k is Boltzmann's constant, and $\Theta$ is the 
characteristic temperature introduced by Debye [32] for the explanation of 
the specific heat of solids. It is $\Theta = h\nu_m$/k, where $\nu_m$ is a 
maximal frequency. In the case of the oscillations making up the 
$\pi^0$\,meson the maximal frequency is $\nu_m = \pi\nu_0$,  therefore 
$\nu_m = 1.5\cdot10^{26}$ sec$^{-1}$, and we find that $\Theta = 
7.2\cdot10^{15}$\,K.
 
  In order to determine T we set the internal energy u equal to 
m$(\pi^0)$c$_\star^2$. It then follows from Eq.(16) that $\Theta$/T = 
30.20, or 
T = 2.38\,$\cdot$\,10$^{14}$\,K. That means that Planck's formula (15) 
introduces a 
factor $1/(e^{\Theta/T} - 1\,) \approx 1/e^{30.2}$ = 1/(1.305$\cdot10^{13}$) 
into Eq.(14). In other words, if we determine the temperature T of the particle 
through 
Eq.(16), and correct Eq.(14) accordingly then we arrive at a sum of the 
oscillation energies of the $\pi^0$\,meson which is 
1.0866$\cdot10^{-4}$\,erg = 67.82\,MeV. That 
means that the sum of the energies of the one-dimensional oscillations 
 consisting of N waves  is 0.502E(m($\pi^0$)). We have to double this
 amount because standing waves 
consist of two waves traveling in opposite direction with the same absolute 
value of the frequency. The sum of the energy of the oscillations 
in the $\pi^0$\,meson is therefore 
\begin{equation} \mathrm{E}_\nu(\pi^0)(theor) = 2.1732\cdot10^{-4}\,\mathrm{erg} = 
135.64\,\mathrm{MeV} = 
1.005\mathrm{E(m}(\pi^0))(exp)\,,\end{equation}
\noindent 
if the oscillations are parallel to the $\phi$ axis. The energy in the 
measured mass of the $\pi^0$\,meson and the energy 
in the sum of the oscillations agree fairly well, considering the 
uncertainties of the parameters involved.  

   To sum up: We find that the energy in the rest mass of the $\pi^0$\,meson
 and the other particles of the $\gamma$-branch are correctly given 
by the sum of the energy of  standing electromagnetic waves in a cube, if 
the energy of the oscillations is determined by Planck's formula for the 
energy of a linear oscillator. The $\pi^0$\,meson is like an adiabatic, 
cubic black body filled with standing electromagnetic waves. We know from 
Bose's work [33] that Planck's formula applies to a photon gas as well.
For all $\gamma$-branch particles we have found a simple mode of 
standing waves in a cubic lattice. Since the equation 
determining the frequency of the standing waves is quadratic it follows 
\emph{automatically} that for each positive frequency there is also a negative 
frequency of the same absolute value, that means that for each particle 
there exists also an antiparticle. For the explanation of the mesons and baryons of 
the $\gamma$-branch $\emph{we use only photons, nothing else}$. A rather 
conservative explanation of the $\pi^0$\,meson and the $\gamma$-branch 
particles which does not use hypothetical particles.

   From the frequency distributions of the standing waves in the lattice 
follow the ratios of the masses of the particles which obey the integer 
multiple rule. It is important to note that in this theory the ratios of 
the masses of the $\gamma$-branch particles to the mass of the 
$\pi^0$\,meson $\emph{do not depend}$ on the sidelength of the lattice, 
and the 
distance between the lattice points, neither do they depend on the 
strength of the force between the lattice points nor on the mass of the 
lattice points. The mass ratios are determined only by the spectra of the 
frequencies of the standing waves in the lattice. 

\section {The neutrino branch particles}

The masses of the neutrino branch, the $\pi^\pm$, K$^{\pm,0}$, n, 
D$^{\pm,0}$ and D$^\pm_s$ particles, are integer multiples of the mass of 
the $\pi^\pm$\,mesons times a factor $0.85\,\pm\,0.02$ as we stated 
before. We assume, 
as appears to be quite natural, that the $\pi^\pm$\,mesons and the \,
 other particles \,of\, the \,$\nu$-branch $\emph{consist of the 
same particles into} \\
\emph{which they decay}$, that means of neutrinos and
 antineutrinos and of electrons or positrons, particles whose existence is
 unquestionable. Since the particles of the $\nu$-branch decay 
through weak decays, we assume, as appears likewise to be natural, that 
\emph{the weak nuclear force holds the particles of the $\nu$-branch 
together}. The existence of the weak nuclear force is also unquestionable. 
Since the range of the weak interaction is only about a thousandth of the 
diameter of the particles, the weak force can hold particles together only 
if the particles have a lattice structure, just as macroscopic crystals are 
 held together by microscopic forces between atoms. In the absence of a
 force which originates in the center of the particle and affects all neutrinos
of the particle the configuration of the particle is not spherical but cubic,
 reflecting the very short range of the weak nuclear force.  We will now 
investigate the energy which is contained in the oscillations of a cubic 
lattice consisting of electron and muon neutrinos and their antiparticles, 
and in their rest masses.

   It will be necessary to outline the basic aspects of diatomic lattice 
oscillations. In $\emph{diatomic}$ lattices the lattice points have 
alternately the masses m and M, as with the masses of the electron
 neutrinos m($\nu_e$) and muon neutrinos m($\nu_\mu$).
 The classic example of a diatomic lattice 
is the salt crystal with the masses of the Na and Cl atoms in the lattice 
points. The theory of diatomic lattice oscillations was started by Born 
and v.\,Karman [14]. They first discussed a diatomic chain. The equation 
of motions in the chain are according to Eq.(22) of B\&K

\begin{equation} \mathrm{m}\ddot{u}_{2n} = \alpha(u_{2n+1} + u_{2n-1} - 2u_{2n})\, 
,\end{equation}

\begin{equation} \mathrm{M}\ddot{u}_{2n+1} = \alpha(u_{2n+2} + u_{2n} - 
2u_{2n+1}) \,, 
\end{equation}
\noindent    
where the u$_n$ are the displacements, n an integer number and $\alpha$ a 
constant characterizing the force between the particles. Eqs.(18,19) are 
solved with

\begin{equation}u_{2n} = Ae^{i(\omega\,t\,+\,2n\phi)} ,\end{equation} 

\begin{equation}u_{2n+1} = 
Be^{i(\omega\,t\,+\,(2n+1)\phi)}\,,\end{equation} 

\noindent
where A and B are constants and $\phi$ is given by $\phi = 
2\pi\emph{a}/\lambda$ as in (6). $\emph{a}$ is the lattice constant as 
before and $\lambda$ the wavelength, $\lambda$ = n$\emph{a}$. The 
solutions of Eqs.(20,21) are obviously periodic in time and space and 
describe again standing waves. Using (20,21) to solve (18,19) leads to a 
secular equation from which according to Eq.(24) of B\&K the frequencies 
of the oscillations of the chain follow from 

\begin{equation}4\pi^2\nu^2_\pm  = \alpha/\mathrm{Mm}\cdot((\mathrm{M+m}) \pm\sqrt{(\mathrm{M-m})^2 + 
4\mathrm{mM}\mathrm{cos}^2\phi}\,)\,.\end{equation}    

Longitudinal and transverse waves are distinguished by the minus or plus 
sign in front of the square root in (22).

\section{The masses of the $\nu$-branch particles}

The characteristic case of the neutrino branch particles are the
 $\pi^\pm$\,mesons which can be created in the process $\gamma$ + p
$\rightarrow \pi^- + \pi^+$ + p. A photon impinges on a proton and is
converted in $10^{-23}$ sec into a pair of particles of opposite charge. Fourier
analysis dictates that a continuum of frequencies must be in the collision
products. The waves must be standing waves in order to be part of the
rest mass of a particle. The $\pi^\pm$\,mesons decay via 
 $\pi^\pm \rightarrow \mu^\pm + \nu_\mu(\bar{\nu}_\mu)$ (99.98770\%)
 followed by e.g. $\mu^+ \rightarrow e^+ + \bar{\nu}_\mu + \nu_e$
 ($\approx$ 100\%). Only $\mu$\,mesons,
 which decay into charge and neutrinos, and neutrinos 
result from the decay of the $\pi^\pm$\,mesons. If the particles
 consist of the particles into which they 
decay, then the $\pi^\pm$\,mesons  are made of neutrinos, antineutrinos
 and e$^\pm$. Since neutrinos interact through the 
weak force which has a range of 10$^{-16}$\,cm according to p.25 of [23], 
and since the size of the nucleon is on the order of 10$^{-13}$\,cm, 
the $\nu$-branch particles must be held together in a lattice. It is not
 known with certainty that neutrinos actually 
have a rest mass as was originally suggested by Bethe [34] and Bahcall 
[35] and what the values of m($\nu_e$) and m($\nu_\mu$) are. However, the 
results of the Super-Kamiokande [36] and the Sudbury [37] experiments 
indicate that the neutrinos have rest masses. The neutrino lattice must be 
diatomic, meaning that the lattice points have alternately larger 
(m($\nu_\mu$)) and smaller (m($\nu_e$)) masses. We will retain the 
traditional term diatomic. \emph{The term neutrino lattice will refer to a lattice
 consisting of neutrinos and antineutrinos}. The lattice we consider is 
shown in Fig.\,2. 
Since the neutrinos have spin 1/2 this is a four-Fermion lattice. The first 
investigation of cubic Fermion lattices in context with the elementary 
particles was made by Wilson [13]. A neutrino lattice is 
electrically neutral. Since we do not know the structure of the electron
 we cannot consider lattices with a charge.

	\begin{figure}[h]
	\hspace{3cm}
	\includegraphics{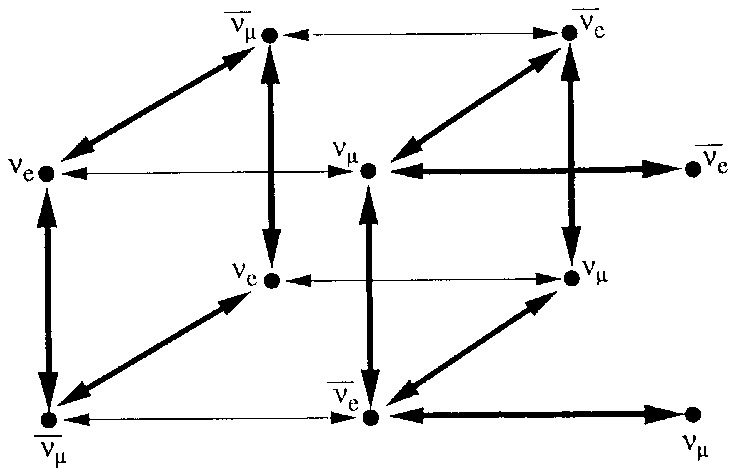}
	\vspace{-0.2cm}
	\begin{quote}
Fig.\,2: A cell in the neutrino lattice. Bold lines mark the forces between 
neutrinos and antineutrinos. Thin lines mark the forces between either 
neutrinos only, or antineutrinos only.
	\end{quote}
	\end{figure}

A neutrino lattice takes care of the continuum of frequencies which must, 
according to Fourier analysis, be present after the high energy collision 
which created the particle. We will, for the sake of simplicity, first set 
the sidelength of the lattice at 10$^{-13}$\,cm that means approximately 
equal to the size of the nucleon. The lattice then contains about 10$^9$ 
lattice 
points, since the lattice constant $\emph{a}$ is on the order of 
10$^{-16}$\,cm. 
The sidelength of the lattice does not enter Eq.(22) 
for the frequencies of the lattice oscillations. The calculation of the 
ratios of the masses  is consequently
 independent of the size of the lattice, as was the case with 
the $\gamma$-branch. The size of the lattice can be explained with 
the pressure which the lattice oscillations exert on a crossection of the 
lattice. The pressure cannot exceed Young's modulus of the lattice. We 
require that the lattice is isotropic.

	\begin{figure}[h]
	\hspace{1.5cm}
	\includegraphics{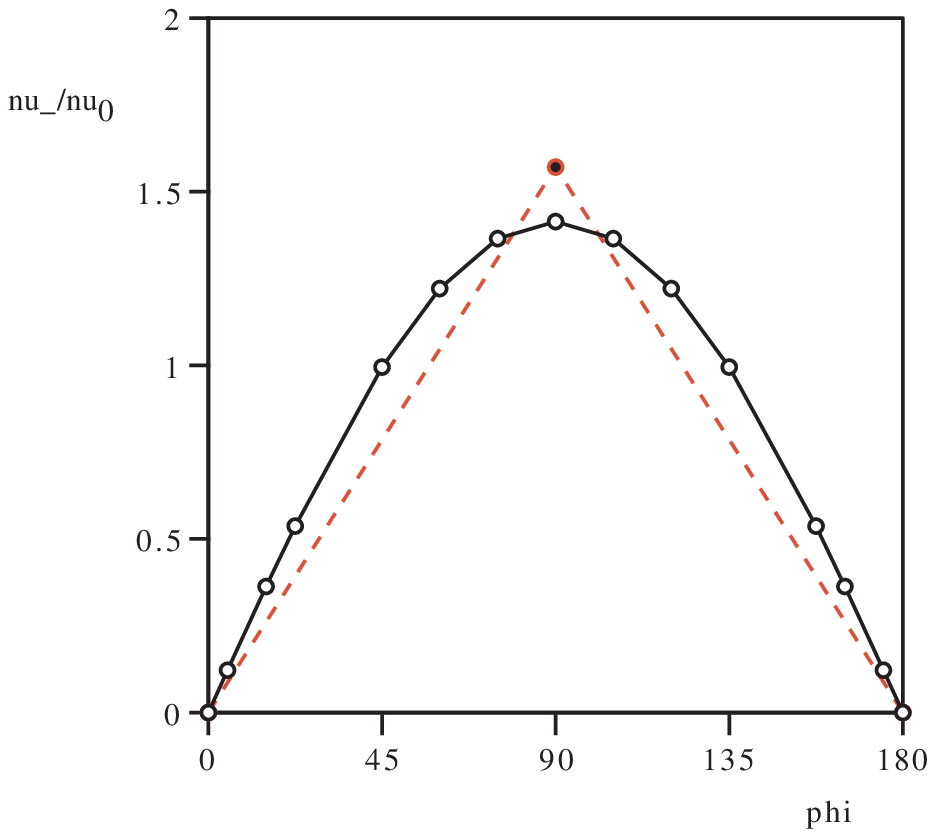}
	\vspace{-.5cm}
	\begin{quote}
Fig.\,3: The frequency distribution $\nu_-/\nu_0$ of the basic diatomic 
mode 
according to Eq.(22) with M/m = 100. The dashed line shows the 
distribution 
of the frequencies corrected for the group velocity limitation.
	\end{quote}
	\end{figure}

   From the frequency distribution of the axial diatomic oscillations 
   (Eq.22), shown in Fig.\,3, follows the group velocity 
   $\mathrm{d}\omega/\mathrm{dk} 
= 2\pi\emph{a}\,\,d\nu/d\phi$\, at each point $\phi$. With $\nu = 
\nu_0f(\phi)$ and $\nu_0$ = $\sqrt{\alpha/4\pi^2\mathrm{M}}$ = 
c$_\star/2\pi\emph{a}$
 as in Eq.(9) we find

\begin{equation} c_g = d\omega/dk = 
\emph{a}\sqrt{\alpha/\mathrm{M}}\cdot df(\phi)/d\phi\,.
\end{equation}
\noindent	
In order to determine the value of d$\omega/dk$ we have to know the value 
of $\sqrt{\alpha/\mathrm{M}}$. From Eq.(8) for $\alpha$ follows that $\alpha = 
\emph{a}\,c_{44}$ in the isotropic case with central forces. The group 
velocity is therefore

\begin{equation} c_g = \sqrt{a^3c_{44}/\mathrm{M}}\cdot df/d\phi\,.
\end{equation}

\noindent
We now set $\emph{a}\sqrt{\alpha/\mathrm{M}}$ = c$_\star$ as in Eq.(11), where 
c$_\star$ is the velocity of light. It follows that

\begin{equation} c_g = c_\star\cdot df/d\phi\,,
\end{equation}

\noindent
as it was with the $\gamma$-branch, only that now on account of 
the rest masses of the neutrinos the group velocity must be smaller than 
c$_\star$, so the value of df/d$\phi$ is limited to $<$\,1, but c$_g 
\cong$\, c$_\star$, 
which is a necessity because the neutrinos in the lattice soon approach 
the velocity of light as we will see. Equation (25) applies regardless 
whether we consider $\nu_+$ or $\nu_-$ in  
Eq.(22). That means that there are no separate transverse 
oscillations with their theoretically higher frequencies.

  The rest mass M of the heavy neutrino can be determined with lattice theory
from Eq.(24) as we have shown in [12]. This involves the inaccurately known
compression modulus of the proton. We will, therefore, rather determine the rest 
mass of the muon neutrino with Eq.(27), which leads to m($\nu_\mu$) $\approx$
50\,milli-eV/c$_\star^2$. It can be verified easily that
 m($\nu_\mu$) = 50\,meV/c$_\star^2$ makes sense. The energy of the 
rest mass of the $\pi^\pm$ mesons is 139\,MeV, and we have N/4 = 
0.7135$\cdot10^9$ 
muon neutrinos and the same number of anti-muon neutrinos. It 
then follows that the energy in the rest masses of all muon and 
anti-muon neutrinos is 
71.35\,MeV, that is 51\% of the energy of the rest mass of the 
$\pi^\pm$ mesons, m($\pi^\pm$)c$_\star^2$ = 139.57\,MeV. A small 
part of m($\pi^\pm$)c$_\star^2$ goes, as we will see, into the electron 
neutrino masses, the rest goes into the lattice oscillations.

   The rest mass of the $\pi^\pm$\,mesons is the 
sum of the oscillation energies and the sum of the rest masses of the 
neutrinos. For the sum of the energies of the frequencies we use Eq.(14) 
with the same N and $\nu_0$ we used for the $\gamma$-branch. For the  
integral in Eq.(14) of the corrected axial diatomic frequencies we find the  
value $\pi^2$/2 as can be easily derived from the plot of the corrected 
frequencies in Fig.\,3. The value of the integral 
 in Eq.(14) for the axial diatomic frequencies $\nu = \nu_0\phi$ is 1/2 of 
the value $\pi^2$ of the same integral in the case of axial monatomic 
frequencies, because in the latter case the increase of the corrected 
frequencies continues to $\phi$ = $\pi$, whereas in the diatomic case the 
increase of the corrected frequencies ends at $\pi$/2, see Fig.\,3.  We 
consider c$_g$ to be so close to c$_\star$ that it does not 
change the value of the integral in Eq.(14) significantly. It can be 
calculated that the time average of 
the velocity of the electron neutrinos in the $\pi^\pm$\,mesons is 
$\bar{v}$ = 0.99994c$_\star$, if m($\nu_e$) = 0.365\,milli-eV/c$_\star^2$
 as will be shown in Eq.(33). Consequently we find that the sum of the
 energies of the corrected diatomic neutrino 
frequencies is 0.5433$\cdot$10$^{-4}$\,erg = 33.91\,MeV. We double 
 this amount because we deal with the superposition of two waves of the 
 same energy and find that \emph{the energy of the neutrino oscillations}
in $\pi^\pm$ is

\begin{equation} \mathrm{E}_\nu(\pi^\pm)  =  67.82\,\mathrm{MeV} = 1/2\cdot\mathrm{E}_\nu(\pi^0) = 0.486\,\mathrm{m}(\pi^\pm)\mathrm{c}_\star^2.
\end{equation}

   In order to determine the sum of the rest masses of the neutrinos we
make use of E$_\nu(\pi^\pm)$ and obtain an approximate value of the
 rest mass of the muon neutrino from 
\begin{equation} \mathrm{m}(\pi^\pm)\mathrm{c}_\star^2 - \mathrm{E}_\nu(\pi^\pm) = \sum\,[m(\nu_\mu) +
 m(\bar{\nu}_\mu) + m(\nu_e) + m(\bar{\nu}_e)]\mathrm{c}_\star^2 = 71.75\,\mathrm{MeV}\,.
\end{equation}
 If m($\nu_e$) $\ll$ m($\nu_\mu$) and m($\nu_\mu$)
 = m($\bar{\nu}_\mu$), as we will justify later, we arrive with N/2 =
 1.427$\cdot10^9$ at

\hspace{4.5cm} m($\nu_\mu$) $\approx$ 50\,milli-eV/$\mathrm{c}_\star^2$.

\noindent The sum of the energy of the rest masses of all neutrinos Eq.(27)
plus the oscillation energy Eq.(26) gives
 \emph{the theoretical rest mass} of the $\pi^\pm$\,mesons which is, since 
we used m($\pi^\pm)$  in the determination of the neutrino rest masses with
Eq.(27), equal to the experimental rest mass of 139.57\,MeV.

   A cubic lattice and conservation of neutrino numbers during the reaction
 $\gamma$  + p $\rightarrow \pi^+ + \pi^- + p$ \emph{necessitates} that
the $\pi^+$ and $\pi^-$  lattices contain just as many  electron
 neutrinos as anti-electron neutrinos. If the lattice has six equal sides there
 must also be a center neutrino in each lattice. Conservation of neutrino numbers
 requires furthermore that the center neutrino of, say, $\pi^+$ is matched by an
 antineutrino in $\pi^-$. However, in the decay sequence of (say)
 the $\pi^-$\,meson $\pi^- \rightarrow  \mu^- + \bar{\nu}_\mu$ and 
$\mu^- \rightarrow$ e$^-$ +\,$\nu_\mu\,+\,\bar{\nu}_e $ an electron
 neutrino does not appear. But since (N\,-\,1)/4 electron 
neutrinos $\nu_e$ must be in the $\pi^-$ lattice (N\,-\,1)/4 electron neutrinos must go
 with the electron emitted in the $\mu^-$ decay. Whether or not this interpretation 
is correct can be decided only after the explanation of the structure of the electron.
  
   The \emph{antiparticle} of the $\pi^+$\,meson is the particle in which all frequencies 
of the neutrino lattice oscillations have been replaced by frequencies with 
the opposite sign, all neutrinos replaced by their antiparticles and the 
positive charge replaced by the negative charge. If, as we will show, the 
antineutrinos have the same rest mass as the neutrinos it follows that the 
antiparticle of the $\pi^+$\,meson has the same mass as $\pi^+$ but 
opposite charge, i.e. is the $\pi^-$\,meson. 
 
   The primary decay of the K$^\pm$\,mesons K$^\pm \rightarrow \mu^\pm + 
\nu_\mu(\bar{\nu}_\mu)$  
(63.5\%), leads to the same end products as the $\pi^\pm$\,meson decay 
$\pi^\pm \rightarrow \mu^\pm + \nu_\mu(\bar{\nu}_\mu)$. From this and the
 composition of the $\mu$\,mesons we learn 
that the K\,mesons must, at least partially, be made of the same four 
neutrino types as in the $\pi^\pm$\,mesons namely of muon neutrinos, 
anti-muon neutrinos, electron neutrinos and anti-electron neutrinos and 
their oscillation energies. However the K$^\pm$\,mesons cannot be 
solely the second mode of the lattice oscillations of the 
$\pi^\pm$\,mesons, because the second mode of the $\pi^\pm$\,mesons 
has an energy of 4E$_{\nu}$($\pi^\pm$) + N/2$\cdot$(m($\nu_\mu$) + 
m($\nu_e$))\,c$_\star^2$ $\approx$ (271.3 + 71.75)MeV = 343\,MeV. The 
343\,MeV 
characterize the second or (2.) mode of the $\pi^\pm$ mesons, which fails 
m(K$^\pm$)c$_\star^2$ = 493.7\,MeV by a wide margin.

   Anyway, the concept of the K$^\pm$\,mesons being alone a higher mode 
of the $\pi^\pm$\,mesons contradicts our point that the particles consist of the 
particles into which they decay. The decays K$^\pm$ $\rightarrow$ 
$\pi^\pm + \pi^0$ (21.13\%), as well as K$^+$ $\rightarrow$ $\pi^0 + 
e^+ + \nu_e$ (4.87\%), called K$^+_{e3}$, and K$^+$ $\rightarrow$ $\pi^0 +
 \mu^+ + 
\nu_\mu$ (3.27\%), called K$^+_{\mu3}$, make up 29.27\% of the K$^\pm$
\,meson decays. A $\pi^0$\,meson figures 
in each of these decays. If we add the energy in the rest mass of a 
$\pi^0$\,meson m($\pi^0$)c$_\star^2$ = 134.97\,MeV to the 343\,MeV in the 
second mode 
of the $\pi^\pm$\,mesons then we arrive at an energy of 478\,MeV, which is 
96.8\% of 
m(K$^\pm$)c$_\star^2$. Therefore we conclude that the K$^\pm$\,mesons 
consist of the second 
 mode of the $\pi^\pm$\,mesons $\emph{plus}$ a $\pi^0$\,meson or
 are the state (2.)$\pi^\pm$ + $\pi^0$. Then 
it is natural that  $\pi^0$\,mesons from the $\pi^0$ component 
in the K$^\pm$\,mesons are among the decay products of the K$^\pm$\,mesons.

   We obtain the K$^0$\,meson if we superpose onto the second mode of the 
$\pi^\pm$\,mesons instead of 
a $\pi^0$\,meson a basic mode of the $\pi^\pm$\,mesons with a charge 
opposite 
to the charge of the second mode of the $\pi^\pm$\,meson. The 
K$^0$ and $\overline{{\mathrm{K}}^0}$\,mesons, 
or the state (2.)$\pi^\pm$ + $\pi^\mp$,  
is made of neutrinos and antineutrinos only, without a photon component,
 because the second mode 
of $\pi^\pm$ as well as the basic mode $\pi^\mp$ consist of neutrinos and
 antineutrinos only.  The K$^0$\,meson has a measured mean square charge
 radius $\langle$r$^2$$\rangle$ = -\,0.076 $\pm$ 0.021\,fm$^2$ according to [38],
which can only be if there are two charges of opposite sign within K$^0$ as our
 model postulates. Since the mass of a $\pi^\pm$\,meson is by
 4.59\,MeV/c$_{\star}^2$ larger
 than the mass of a $\pi^0$\,meson the mass of K$^0$ should be larger than 
m(K$^\pm$), and indeed m(K$^0$)\,$-$\,m(K$^\pm$) = 3.995\,MeV/c$_\star^2$ 
according to [2]. Similar differences occur with m(D$^\pm$)\,$-$\,m(D$^0$) and 
m($\Xi_c^0$)\,$-$\,m($\Xi_c^+$). The 
decay K$^0_S \rightarrow \pi^++ \pi^-$ (68.6\%) creates directly the 
$\pi^+$ and $\pi^-$ mesons which are part of the (2.)$\pi^\pm$ + 
$\pi^\mp$ structure of K$^0$ we have 
suggested. The decay K$^0_S \rightarrow \pi^0 + \pi^0$ (31.4\%) apparently 
originates from the 2$\gamma$ branch of electron positron annihilation. 
Both decays account for 100\% of the decays of K$^0_S$. 
The decay K$^0_L \rightarrow 3\pi^0$ (21.1\%) apparently comes from the 
3$\gamma$ branch of electron positron annihilation. The two decays of 
K$^0_ L$ called K$^0_{\mu3}$ into $\pi^\pm\,\mu^\mp\,\nu_\mu$ 
(27.18\%) and K$^0_{e3}$ into $\pi^\pm\,e^\mp\,\nu_e$ (38.79\%) 
which together make up 65.95\% of 
the K$^0_{L}$ decays apparently originate from the decay of the 
second mode of the $\pi^\pm$\,mesons in the 
K$^0$ structure, either into $\mu^\mp$ + $\nu_\mu$  
 or into $e^\mp$ + $\nu_e$. The same types of decay, 
 apparently tied to the (2.)$\pi^\pm$ mode, accompany also the K$^\pm$ 
 decays where, however, a $\pi^0$\,meson replaces the $\pi^\pm$\,mesons in 
 the K$^0_L$ decay products. Our rule 
that the particles consist of the particles into which they decay also 
holds for the K$^0$ and $\overline{{\mathrm{K}}^0}$\,mesons. The
explanation of the K$^0$,$\overline{{\mathrm{K}}^0}$\,mesons with the
state (2.)$\pi^\pm$ + $\pi^\mp$ confirms that the state (2.)$\pi^\pm$ +
$\pi^0$ was the correct choice for the explanation of the K$^\pm$\,mesons.
The state (2.)$\pi^\pm$ + $\pi^\mp$ is also crucial for the explanation of the 
absence of spin of the K$^0$,$\overline{{\mathrm{K}}^0}$\,mesons, as we will
see later.   
   
   The neutron with a mass $\approx2\mathrm{m}(\mathrm{K}^\pm$) is either the 
superposition of a K$^+$ and a K$^-$\,meson or of a  
K$^0$\,meson and a $\overline{{\mathrm{K}}^0}$\,meson. As
 we will see, the spin rules out
  a neutron consisting of a K$^+$ and a K$^-$\,meson. On the other hand, the
 neutron can consist of a K$^0$ and a 
$\overline{{\mathrm{K}}^0}$\,meson.
 In this case the neutron lattice contains at each lattice point a 
$\nu_\mu,\bar{\nu}_\mu,\nu_e,\bar{\nu}_e$ neutrino quadrupole plus the
 second mode of the  
lattice  oscillations and a quadrupole of positive and negative electrical charges.
 The lattice oscillations in the neutron must be coupled  in 
order for the neutron to have spin 1/2, just as the $\Lambda$ 
baryon with spin 1/2 is a superposition of two $\eta$\,mesons. With 
m(K$^0$)(theor) = m(K$^\pm$) + 4\,MeV/c$^2$ = 482\,MeV/c$^2_\star$
 from above it follows that m(n)(theor) $\approx$ 
2m(K$^0)$(theor) $\approx$ 964\,MeV/c$^2_\star$ = 1.026m(n)(exp). 

   The proton does not decay and does not tell which particles it is 
made of. However, we learn about the structure of the proton through the 
decay of the neutron n $\rightarrow$ p + e$^- + \bar{\nu}_e$ (100\%). 
One single anti-electron neutrino is emitted when the neutron decays and 
1.293 MeV are released. But there 
is no place for a permanent vacancy of a single missing neutrino and for a small 
amount of permanently missing oscillation energy in a nuclear lattice. As 
it appears all anti-electron neutrinos are removed from the structure of 
the neutron in the neutron decay and converted into
 the kinetic energy of the decay products. This type of 
process will be discussed again in the following 
section. On the other hand, it seems to be certain that 
the proton consists of a neutrino lattice carrying a net positive electric charge.
Actually, in our model the proton contains three charges e$^+$e$^-$e$^+$.
The concept that the proton carries just one electrical charge has been
 abandoned a long time ago when it was said that the proton consists of three
 quarks carrying fractional electrical charges. 
Each elementary charge in the proton has a magnetic moment,
 all of them point in the same direction because
 the spin of the one e$^-$ must be opposite to the spin of the two e$^+$.
Each magnetic moment of the elementary charges has a g-factor $\cong$\,2.
All three electric charges in the proton must then have a g-factor $\approx$\,6,
whereas the measured g(p) = 5.585 = 0.93$\cdot$6.   

 The D$^\pm$\,mesons with m(D$^\pm$) = 0.9954\,(m(p) + 
m($\bar{\mathrm{n}}$))
are the superposition of a proton and an antineutron of opposite spin or 
of their antiparticles, whereas the 
superposition of a proton and a neutron with the same spin creates the 
deuteron with spin 1 and a mass m(d) = 0.9988$\cdot$(m(p) + m(n)). In this 
case the proton and neutron interact with the strong force, nevertheless 
the deuteron consists of a neutrino lattice with standing waves. The 
D$_s^\pm$\,mesons seem 
to be made of a body centered cubic lattice as discussed in [39]. 

   The average 
factor 0.85 $\pm$ 0.025 in the ratios of the particles of the $\nu$-branch 
to the $\pi^\pm$\,mesons is a consequence of the neutrino rest masses. 
They 
make it impossible that the ratios of the particle masses are integer 
multiples because the 
particles consist of the energy in the neutrino oscillations plus the 
neutrino rest masses which are independent of the order of the lattice 
oscillations. Since the contribution in percent of the neutrino rest 
masses to the $\nu$-branch particle masses decreases with increased 
particle 
mass the factor in front of the mass ratios of the $\nu$-branch particles 
must decrease with increased particle mass.

   Summing up: The characteristic feature of the $\nu$-branch particles is the 
cubic lattice consisting of $\nu_\mu,\bar{\nu}_\mu,\nu_e,\bar{\nu}_e$ neutrinos.
The rest masses of the $\nu$-branch particles comes from the sum of the rest
 masses of the neutrinos and antineutrinos and from the energy in the lattice
 oscillations. The existence of
 the neutrino lattice is a necessity if one wants to explain the spin, or the absence
 of spin, of the $\nu$-branch particles. For the explanation of the $\nu$-branch
 particles we do not use hypothetical particles either.   

\section{The rest masses of the leptons}

Surprisingly one can also explain the mass of the $\mu^\pm$\,mesons with 
the standing wave model. The $\mu$\,mesons are part of 
the lepton family which is distinguished from the mesons and baryons by the 
absence of strong interaction with the mesons and baryons. The leptons make
 up 1/2 of the number of stable elementary particles. The standard model of
 the particles does not deal with the lepton masses. Barut [40] has given a simple
 and quite accurate empirical formula relating the masses of 
the electron, $\mu$\,meson and $\tau$\,meson, which formula has been 
extended by Gsponer and Hurni [41] to the quark masses.

   The mass of the $\mu$\,mesons is m($\mu^\pm$) = 105.658357 $\pm$ 
5$\cdot$$10^{-6}$\,MeV/c$_\star^2$, according to the Review of Particle
 Physics [2]. The mass of the $\mu$\,mesons is usually compared to the mass 
of the electron and is often said to be m($\mu$) = m(e)(1 + 
3/2$\alpha_f$) = 206.554\,m(e), ($\alpha_f$ being the fine structure 
constant), whereas the experimental value is 
206.768\,m(e). The $\mu$\,mesons are ``stable", their lifetime is 
$\tau(\mu^\pm) 
= 2.19703\cdot10^{-6} \pm 4\cdot10^{-11}$\,sec, about a hundred times 
longer than the lifetime of the $\pi^\pm$\,mesons, that means longer than 
the lifetime of any other elementary particle, but for the electrons, protons 
and neutrons.

   Comparing the mass of the $\mu$\,mesons to the mass of the 
$\pi^\pm$\,mesons 
m($\pi^\pm$) = 139.57018\,MeV/c$_\star^2$ we find that 
m($\mu^\pm$)/m($\pi^\pm$) = 0.757027 = 1.00937\,$\cdot$\,3/4 or that 
that the mass of the $\mu^\pm$ mesons is in a good approximation 3/4 
of the mass of the $\pi^\pm$\,mesons. We have also m($\pi^\pm)\,-$ 
m($\mu^\pm$) = 33.9118\,MeV/c$_\star^2$ = 0.24297m($\pi^\pm$) or 
approximately 1/4\,$\cdot$\,m$(\pi^\pm$).
 The mass of the electron is approximately 1/206 of the mass of the muon,
 the contribution of m(e$^\pm$) to m($\mu^\pm$) will therefore 
be neglected in the following. We assume, as we have done before and as 
appears to be natural, that the particles, including the muons, 
$\emph{consist of the particles into}$ \emph{which they decay}. The 
$\mu^+$\,meson 
decays via $\mu^+ \rightarrow e^+ + \bar{\nu}_\mu + \nu_e$ ($\approx$ 
100\%). The muons are apparently composed of some of the neutrinos,
antineutrinos and their oscillations which 
are present in the cubic neutrino lattice of the $\pi^\pm$\,mesons 
according to our standing wave model.

   In the standing wave model the  
$\pi^\pm$\,mesons are composed of a cubic lattice consisting of N = 
2.854\,$\cdot\,10^9$  neutrinos and antineutrinos. We must now be
 more specific about N, which is an odd number, because a lattice with six
sides of equal composition has a center particle, just as the NaCl lattice.
In the $\pi^\pm$\,mesons there are then (N\,-\,1)/4 muon neutrinos 
$\nu_\mu$ and the same number of 
anti-muon neutrinos $\bar{\nu}_\mu$,  
as well as (N\,-\,1)/4 electron neutrinos $\nu_e$ and the same number of 
anti-electron neutrinos $\bar{\nu}_e$, plus a center neutrino or antineutrino.
We replace N\,-\,1 by N$^\prime$. Since N$^\prime$ differs from N by 
 only one in $10^9$ we have N$^\prime$ $\cong$ N. Although the numerical
 difference between N and N$^\prime$ is negligible we cannot consider  
e.g. N/4 neutrinos because that would mean that there would be fractions of a
 neutrino. N$^\prime$ is an integer multiple of 4, because of the equal numbers
of $\nu_e,\bar{\nu}_e,\nu_\mu,\bar{\nu}_\mu$ neutrinos.  

   From Eq.(27) followed that the rest mass of a muon neutrino should be 
about 50\,milli-eV/c$_\star^2$. Provided that 
the mass of an electron neutrino is small as compared to 
m($\nu_\mu$) we find, with N = 2.854$\cdot10^9$, that:
\bigskip

(a) The difference of the rest masses of the $\mu^\pm$ and $\pi^\pm$ 
mesons is nearly equal to the sum of the rest masses of all muon, 
respectively anti-muon, neutrinos in the $\pi^\pm$\,mesons.
\bigskip

\noindent
    m($\pi^\pm)\,-\,$m$(\mu^\pm$) = 33.912\,MeV/$\mathrm{c}_\star^2$ \quad versus
\quad $\mathrm{N^\prime}/4\cdot\mathrm{m}(\nu_\mu)$ $\approx$
 35.68\,$\mathrm{MeV}/\mathrm{c}_\star^2$\,.

\bigskip

(b) The energy in the oscillations of all $\nu_\mu,\bar{\nu}_\mu,\nu_e,\bar{\nu}_e$
 neutrinos in the $\pi^\pm$\,mesons
 is nearly the same as the energy in the oscillations of 
all $\bar{\nu}_\mu,\nu_e,\bar{\nu}_e$, respectively 
$\nu_\mu,\nu_e,\bar{\nu}_e$,    neutrinos in the $\mu^\pm$\,mesons.
The oscillation energy is the rest mass of a particle minus the sum of the
 rest masses of all neutrinos in the particle as in Eq.(27). So
\begin{equation}    \mathrm{E}_{\nu}(\pi^\pm) = \mathrm{m}(\pi^\pm)\mathrm{c}_\star^2 - 
\mathrm{N^\prime}/2\cdot[\mathrm{m}(\nu_\mu) + 
\mathrm{m}(\nu_e)]\mathrm{c}_\star^2 = 68.22\,\mathrm{MeV} \end{equation}
\quad versus 
\begin{equation} \mathrm{E}_{\nu}(\mu^\pm) = \mathrm{m}(\mu^\pm)\mathrm{c}_\star^2 - 
\mathrm{N^\prime/4}\cdot \mathrm{m}(\nu_\mu)\mathrm{c}_\star^2 - 
\mathrm{N^\prime}/2\cdot\mathrm{m}(\nu_e)\mathrm{c}_\star^2 = 
69.98\,\mathrm{MeV}\,.\end{equation}

\noindent
(a) seems to say that the energy of the rest masses of all muon (respectively
 anti-muon) neutrinos of the lattice is consumed in the $\pi^\pm$ decay. We attribute 
the 1.768\,MeV difference  between the left and right side of (a) to the second
order effects which cause the deviations of the masses of the particles from
the integer multiple rule. There is also the difference that the left side of (a) deals
with two charged particles, whereas the right side deals with neutral particles.
(b) seems to say that
the oscillation energy of all neutrinos in the $\pi^\pm$ lattice is conserved
in the $\pi^\pm$ decay, which seems to be necessary because the 
oscillation frequencies in $\pi^\pm$ and $\mu^\pm$ must follow Eq.(13)
as dictated by the group velocity limitation. If indeed 
\begin{equation} \mathrm{E}_\nu(\pi^\pm) = \mathrm{E}_\nu(\mu^\pm)
\end{equation} 
then it follows from the difference of Eqs.(28) and (29) that
 m($\pi^\pm$)\,$-$\,m($\mu^\pm$) = N$^\prime$/4$\cdot$m($\nu_\mu$).

   We should note that in the $\pi^\pm$ decays only \emph{one single} 
 muon neutrino is emitted, not N$^\prime$/4 of them, but that in the 
 $\pi^\pm$ decay 33.9\,MeV are released. Since according to (b) the 
oscillation energy of the neutrinos in the $\pi^\pm$ mesons is conserved 
in their decay the 33.9\,MeV released in the $\pi^\pm$ decay can come 
from \emph{no other source} then from the rest masses of the
 muon or anti-muon neutrinos in the $\pi^\pm$\,mesons. The energy in the rest 
 masses of these muon neutrinos is used to supply the kinetic  
energy in the momentum of the single emitted muon neutrino (pc$_\star$ = 
30\,MeV) and in 
the momentum of the emitted $\mu$\,meson (pc$_\star$ = 4\,MeV). The 
average energy of the neutrinos in the $\pi^\pm$ lattice is about
50\,milli-eV, so it is not possible for a single neutrino in $\pi^\pm$ to possess
an energy of 33.9\,MeV. The 33.9\,MeV can come only from the sum of the
muon neutrino rest masses. However, what happens then to the neutrino
 numbers? Either conservation of neutrino numbers is violated or the decay
 energy comes from equal numbers  of muon and anti-muon neutrinos. Equal
numbers N$^\prime$/8  muon and anti-muon neutrinos would then be in
the $\mu^\pm$\,mesons. This would not make a difference in either the 
oscillation energy or in the sum of the rest masses of the neutrinos or in the
spin of the $\mu^\pm$\,mesons. The 
sum of the spin vectors of the N$^\prime$/4 muon or anti-muon neutrinos
 converted into kinetic energy is zero, as will become clear in Section 9.

   Inserting m($\pi^\pm$)\,$\mathrm{-}$\,m($\mu^\pm$) = N$^\prime$/4$\cdot$m($\nu_\mu$)
from (a) into Eq.(28) we arrive at an equation for \emph{the 
theoretical value of the mass of the $\mu^\pm$\,mesons}. It is
\begin{equation} \mathrm{m}(\mu^\pm)\mathrm{c}_\star^2 = 1/2\cdot[\,\mathrm{E}_\nu(\pi^\pm) + 
\mathrm{m}(\pi^\pm)\mathrm{c}_\star^2 +
\mathrm{N^\prime\,m}(\nu_e)\mathrm{c}_\star^2/2\,] = 103.95\,\mathrm{MeV}\,\,,\end{equation}
which expresses m($\mu^\pm$) through the well-known mass of $\pi^\pm$,
the calculated oscillation energy of $\pi^\pm$, and a small contribution (0.4\%)
of the electron neutrino and anti-electron neutrino masses.
A different form of this equation is, with E$_\nu(\pi^\pm$) = E$_\nu(\mu^\pm$),
\begin{equation}\mathrm{m}(\mu^+) = \mathrm{E}_\nu(\mu^\pm)/\mathrm{c}_\star^2 + 
\mathrm{N^\prime m}(\nu_\mu)/4 + \mathrm{N^\prime}[\,\mathrm{m}(\nu_e) + 
\mathrm{m}(\bar{\nu}_e)\,]/4\,\,. \end{equation}
Eq.(31) shows that our explanation of the mass of the $\mu^\pm$\,mesons 
comes close to the experimental value m($\mu^\pm$) = 105.658\,MeV/c$_\star^2$.

   Our model of the $\mu^\pm$\,mesons means that the 
 $\mu$\,mesons have the same size as the $\pi^\pm$\,mesons, 
 namely 0.88$\cdot10^{-13}$ cm. This contradicts 
the commonly held belief that the $\mu$\,mesons are \emph{point particles}. 
However, since in our model the $\mu$\,mesons consist of a neutrino lattice 
plus an electric charge and since neutrinos do not interact, in a good 
approximation, with charge or mass it will not be possible to 
determine the size of the $\mu$\,meson lattice through conventional 
scattering experiments. Therefore the $\mu$\,mesons will appear to be 
point particles.

   Finally we must address the question for what reason do the muons or
 leptons not interact strongly with the 
mesons and baryons? We have shown in [9] that a strong force 
emanates from the sides of a cubic lattice caused by the unsaturated weak 
forces of about $10^6$  lattice points at the surface of the lattice of 
the mesons and baryons. This follows from the study of Born and Stern [42] 
which dealt with the forces between two parts of a cubic lattice cleaved 
in vacuum. If the muons have a lattice consisting of one type of muon neutrinos,
 say, $\bar{\nu}_\mu$ and of $\nu_e$ and $\bar{\nu}_e$ 
neutrinos their octahedronal lattice surface is not the same as the 
surface of the cubic 
$\nu_\mu,\,\bar{\nu}_\mu,\,\nu_e,\,\bar{\nu}_e$ lattice of the mesons and 
baryons in the standing wave model. Therefore  the muon lattice does not 
bond with the cubic lattice of the mesons and baryons.

   To summarize what we have learned about the $\mu^\pm$\,mesons.
Eq.(32) says that the energy in m($\mu^\pm$)c$_\star^2$ is the sum of the oscillation
energy plus the sum of the energy of the rest masses of the neutrinos and
 antineutrinos in m($\mu^\pm$), similar to the energy in the $\pi^\pm$\,mesons.
 The three neutrino types in the $\mu^\pm$\,mesons are the
 remains of the cubic neutrino lattice in the $\pi^\pm$\,mesons. Since all $\nu_\mu$
respectively all $\bar{\nu}_\mu$ neutrinos have been removed from the $\pi^\pm$
lattice in the $\pi^\pm$ decay the rest mass of the $\mu^\pm$\,mesons must be
  $\cong$ 3/4$\cdot$m($\pi^\pm$),  in agreement with the experimental results. 
The $\mu^\pm$\,mesons are not point particles.  

   The mass of the $\tau^\pm$\,mesons follows from the decay of the
 D$^\pm_s$\,mesons. It can be shown readily that the oscillation energies of the
 lattices in
 D$^\pm_s$ and in $\tau^\pm$ are the same. From that follows that the energy
 in the rest mass of the $\tau^\pm$\,mesons is the sum of the oscillation energy in
the $\tau$\,meson lattice plus the sum of the energy of the rest masses of all 
neutrinos and antineutrinos in the $\tau$\,meson lattice, just as with the 
$\mu^\pm$\,mesons. We will skip the details. The tau mesons are not point
 particles either.

   If the same principle that applies to the decay 
of the $\pi^\pm$\,mesons, namely that in the decay the oscillation energy
 of the decaying particle is conserved and that an entire neutrino type 
supplies the energy released in the decay, also applies to the decay of the 
neutron $n\,\rightarrow\,p\,+\,e^-\,+\,\bar{\nu}_e$, then the mass of the 
anti-electron neutrino can be determined from the known difference 
$\Delta$ =  m(n)\,$-$\,m(p) = 1.293332\,MeV/c$_\star^2$. Nearly
 one half of $\Delta$ comes 
from the energy lost by the emission of the electron, whose mass is 
0.510999\,MeV/c$_\star^2$. N anti-electron neutrinos are in the neutrino
 quadrupoles in the neutron, 
one-fourth of them is carried away by the emitted electron. We have seen in
 the paragraph below Eq.(27) that the decay sequence of the $\pi^\pm$\,mesons
requires that the electron carries with it N$^\prime$/4 electron neutrinos, if the
 $\pi^\pm$\,mesons consist of a lattice with a center neutrino or antineutrino  
and equal numbers of
 $\nu_e,\bar{\nu}_e,\nu_\mu,\bar{\nu}_\mu$ neutrinos as required by conservation 
of neutrino number during the creation of $\pi^\pm$.  The electron can carry
 N$^\prime$/4 anti-electron neutrinos as well. Since, as we will see shortly,
 m$(\nu_e$) = m($\bar{\nu}_e$) this does not make a difference energetically
 but is relevant for the orientation of the spin vector of the emitted electron.
After the neutron has lost N$^\prime$/4 anti-electron neutrinos to the electron 
emitted in the $\beta$-decay the remaining 3/4$\cdot$N$^\prime$
 anti-electron neutrinos in the neutron provide
 the energy $\Delta$ $-$ m($e^-$)c$_\star^2$ = 
0.782321\,MeV released in the decay of the neutron. After division by
 3/4$\cdot$N$^\prime$ the rest mass of the anti-electron neutrino is
\begin {equation} \mathrm{m}(\bar{\nu}_e) = 
0.365\,\mathrm{meV/c}_\star^2\,.\end{equation}

\noindent
 Since theoretically the antineutron decays as 
$\bar{n} \rightarrow\bar{p} + e^++\nu_e$ it follows from the same 
considerations as with the decay of the neutron that 
\begin{equation} \mathrm{m}(\nu_e) = \mathrm{m}(\bar{\nu}_e)\,. \end{equation}
We note that 
\begin{equation}\mathrm{N^\prime}/4\cdot \mathrm{m}(\nu_e) =
 \mathrm{N^\prime}/4\cdot \mathrm{m}(\bar{\nu}_e) =
 0.51\,\mathrm{m}(e^\pm)\,. \end{equation}
 
   Inserting (34) into Eq.(27) we find that 

\begin{equation} \mathrm{m}(\nu_\mu) = 49.91\,\mathrm{meV}/\mathrm{c}_\star^2\,,
\end{equation}
Since the same considerations apply for either the $\pi^+$ or 
the $\pi^-$ meson it follows that  
\begin{equation}\mathrm{m}(\nu_\mu) = \mathrm{m}(\bar{\nu}_\mu)\,.
\end{equation}
Experimental values for the rest masses of the different neutrino types are
not available. However it appears that for the $\nu_\mu \leftrightarrow \nu_e$
oscillation the value for $\Delta$m$^2$ = m$^2_2$ $\mathrm{-}$ m$^2_1$  =
3.2$\times10^{-3}$\,eV$^2$ given on p.1565 of [36] can be used to
 determine m$_2$ = m($\nu_\mu$) if m$_1$ = m$(\nu_e$) is much smaller than
 m$_2$. We have then m($\nu_\mu$) $\approx$ 56.56\,milli-eV/c$^2_\star$,
which is compatible with the value of m($\nu_\mu$) given in Eq.(36).

   The mass of the $\tau$\,neutrino can be determined from the decay 
D$_s^\pm$  $\rightarrow$ $\tau^\pm$\,+\,$\nu_\tau\,(\bar{\nu}_\tau)$, and 
 the subsequent decay $\tau^\pm \rightarrow \pi^\pm + \bar{\nu}_\tau
(\nu_\tau)$, as discussed in [39]. The appearance of the $\tau$\,meson
in the decay of D$^\pm_s$ 
and the presence of $\nu_\mu,\bar{\nu}_\mu,\nu_e,\bar{\nu}_e$ neutrinos 
in the $\pi^\pm$ decay product of the $\tau^\pm$\,mesons means that there
 must be $\nu_\tau,\bar{\nu}_\tau,\nu_\mu,\bar{\nu}_\mu,\nu_e,\bar{\nu}_e$ 
neutrinos in the D$^\pm_s$ lattice. The additional $\nu_\tau$ and $\bar{\nu}_\tau$
neutrinos can be accomodated in the D$^\pm_s$ lattice by a body-centered
 cubic lattice, in which there is in the center of each cubic cell one particle different
from the particles in the eight cell corners. In a body-centered cubic lattice there
are N$^\prime$/4  cell centers, if N$^\prime$ is the number of lattice points
 without the cell centers. If 
the center particles are tau neutrinos there must be N$^\prime$/8 $\nu_\tau$ and
N$^\prime$/8 $\bar{\nu}_\tau$ neutrinos, because of conservation of neutrino
 numbers. From m(D$^\pm_s$) = 1968.5\,MeV/c$_\star^2$ and m($\tau^\pm$) = 
1777\,MeV/c$_\star^2$  follows that

\begin{equation} \mathrm{m}(\mathrm{D}^\pm_s) - \mathrm{m}(\tau^\pm) = 191.5\,\mathrm{MeV/c_\star^2} = \mathrm{N^\prime}/8\cdot\mathrm{m}(\nu_\tau)\,.
\end{equation}
\noindent
The rest mass of the $\tau$\,neutrinos is therefore
\begin{equation} \mathrm{m}(\nu_\tau) = \mathrm{m}(\bar{\nu}_\tau) = 0.537\,\mathrm{eV/c}_\star^2\,.
\end{equation}

    \emph{We can now explain the ratio m($\mu^\pm$)/m(e$^\pm$)} as well as 
 m($\pi^\pm$)/m(e$^\pm$) and m(p)/m(e$^-$). From Eqs.(33,36) we find that 

\begin{equation} \mathrm{m}(\nu_\mu)/\mathrm{m}(\nu_e) = 136.74\,,
\end{equation}

\noindent
which is 99.8\% of the inverse of the fine structure constant $\alpha_f$ = 1/137.036.
 It does not seem likely that this is just a coincidence. 
We set N$^\prime$/4$\cdot$m($\nu_e$)
= 0.5\,m(e$^\pm$), not at 0.51\,m(e$^\pm$) as in Eq.(35). Then m(e$^\pm$)
= N$^\prime$/2$\cdot$m($\nu_e$) or N$^\prime$/2$\cdot$m($\bar{\nu}_e$).
 We also set E$_\nu(\pi^\pm$) = 0.5\,m($\pi^\pm$)c$_\star^2$, not at 0.486\,m($\pi^\pm$)c$_\star^2$
as in Eq.(26). With E$_\nu(\pi^\pm$) = E$_\nu(\mu^\pm$) from Eq.(30)
it follows with Eq.(28) that E$_\nu(\mu^\pm$) = 
N$^\prime$/2$\cdot$[m($\nu_\mu)$ + m($\nu_e$)]c$_\star^2$. 
From Eq.(32) then follows that 

\begin{equation} \mathrm{m}(\mu^\pm) = 3/4\cdot\mathrm{N^\prime}
\mathrm{m}(\nu_\mu) + \mathrm{N^\prime}\mathrm{m}(\nu_e)\,,
\end{equation}
and with m(e$^\pm$) = N$^\prime$/2$\cdot$m($\nu_e$) from
above we have 

\begin{equation} \frac{\mathrm{m}(\mu^\pm)}{\mathrm{m(e}^\pm)} = 
\frac{3}{2}\cdot\frac{\mathrm{m}(\nu_\mu)}{\mathrm{m}(\nu_e)} + 2\,,
\end{equation}
or with m($\nu_\mu$)/m($\nu_e$) $\cong$ 1/$\alpha_f$ it turns out that

\begin{equation} \frac{\mathrm{m}(\mu^\pm)}{\mathrm{m(e}^\pm)}
\cong \frac{3}{2}\cdot \frac{1}{\alpha_f} + 2 = 207.55\,. \end{equation}

   The mass of the muon is, according to Eq.(42), much larger than the mass
 of the electron because
the mass of the muon neutrino, which is dominant in the muon, is so much larger 
(m($\nu_\mu)\alpha_f$ $\cong$ m($\nu_e$)) than the mass of the electron
 neutrino. The ratio of the mass of the muon to the mass of the electron is
 independent of the number N$^\prime$ of the neutrinos of either type in both
 lattices. The empirical formula for the mass ratio is m($\mu^\pm$)/m(e$^\pm$) = 
3/2$\alpha_f$ + 1 = 206.55, whereas the actual ratio is 206.768. The empirical
 formula was given by Barut [43], following an earlier suggestion by Nambu [10]
 that m($\mu$) $\approx$ m(e)$\cdot$3/2$\alpha$. We attribute the 0.5\% 
difference between Eq.(43) and the empirical 
formula for m($\mu^\pm$)/m(e$^\pm$) to the absence of electrical charge 
on the right hand side of Eq.(42). The $\mu^\pm$\,mesons, which have spin 
and a magnetic moment, have +\,1 added to the ratio m($\nu_\mu$)/m($\nu_e$)
in Barut's formula for m($\mu^\pm$)/m(e$^\pm$), whereas the 
$\pi^\pm$\,mesons, which do not have spin and a magnetic moment, have 
$\mathrm{-}$1 subtracted from m($\nu_\mu$)/m($\nu_e$) in the formula for
m($\pi^\pm$)/m(e$^\pm$) given in the following. 

Similarly we obtain 
\begin{equation} \frac{\mathrm{m}(\pi^\pm)}{\mathrm{m(e^\pm)}}
= 2\,[\frac{\mathrm{m}(\nu_\mu)}{\mathrm{m}(\nu_e)} +1] 
 \cong \frac{2}{\alpha_f} + 2 = 276.07\,,
\end{equation}
whereas the empirical formula is m($\pi^\pm$)/m(e$^\pm$) = 2/$\alpha_f$\,$-$\,1
= 273.07. The experimental ratio is m($\pi^\pm$)/m(e$^\pm$) = 273.132 =
1.00022\,(2/$\alpha_f\, \mathrm{-}\,1$).

   In order to determine m(n)/m(e$^\pm$) we start with K$^0$ = (2.)$\pi^\pm$
+ $\pi^\mp$ and (2.)$\pi^\pm$ = 4E$_\nu(\pi^\pm$) + N$^\prime$/2$\cdot$[m($\nu_\mu)$ + m$(\nu_e)$]c$_\star^2$.
Then m(K$^0$) =  7N$^\prime$/2$\cdot$[m($\nu_\mu)$ + m($\nu_e$)],
 and with m(n) = 0.9439$\cdot$2\,m(K$^0$) it follows that

\begin{equation} \frac{\mathrm{m(n)}}{\mathrm{m(e^\pm)}} = 0.9439\cdot14\,[\frac{\mathrm{m}(\nu_\mu)}
{\mathrm{m}(\nu_e)} + 1] = 1824.2\,,
\end{equation}
that is 99.2\% of the experimental value 1838.68. With m(p)
= 0.9986$\cdot$m(n) we have 
\begin{equation} \frac{\mathrm{m(p)}}{\mathrm{m(e^\pm)}} =
 0.9426\cdot14\,[\frac{\mathrm{m}(\nu_\mu)}{\mathrm{m}(\nu_e)} + 1] \cong
0.9426\,[\frac{14}{\alpha_f} + 14] = 1821.5\,,
\end{equation}
that is 99.2\% of the experimental value 1836.16. The empirical formula for
m(p)/m(e$^\pm$) is m(p)/m(e$^\pm$) = 14\,[1/$\alpha_f$ $\mathrm{-}$ 6] =
0.9991\,m(p)/m(e$^\pm$)(exp).
 
   To summarize what we have learned about the masses of the leptons: We have
found an explanation for the mass of the $\mu^\pm$\,mesons and $\tau^\pm$\,mesons.
We have also determined the rest masses of the \emph{e},\,$\mu,\tau$ neutrinos and 
antineutrinos.
In other words, we have found the masses of all leptons, exempting the electron, 
a topic which we will deal with later.

\section{The spin of the $\gamma$-branch particles}

  It appears to be crucial for the validity of a model of the elementary 
particles that the model can also explain the spin of the particles 
without additional assumptions. The spin or the intrinsic angular momentum 
is, after the mass, the 
second most important property of the elementary particles. As is 
well-known the spin 
of the electron was discovered by Uhlenbeck and Goudsmit [44] more than 75 
years ago. Later on it was established that the baryons have spin as well, 
but not the mesons. We have proposed an explanation of the spin of the 
particles in [45].  For current efforts to 
understand the spin of the nucleon see Jaffe [46] and of the 
spin structure of the $\Lambda$ baryon see G\"ockeler et al. [47]. The
 explanation of the spin requires an unambiguous answer, the spin must
 be 0 or 1/2 or integer multiples thereof, nothing else. 

    The spin of the particles is, of course, the sum of the angular 
momentum vectors of the oscillations plus the sum of the spin vectors of
 the  neutrinos, antineutrinos and the 
electric charges in the cubic lattice of the mesons and baryons. It is
 striking that the particles which, according to the standing wave model, 
consist of a single oscillation mode do not have spin, as the $\pi^0, 
\pi^\pm$ and $\eta$ mesons do, see Tables 1 and 2. 
 It is also striking that particles whose mass is approximately 
twice the mass of a smaller particle have spin 1/2 as is the 
case with the $\Lambda$ baryon, m($\Lambda$) $\approx$ 2m($\eta$), and 
with the 
nucleon m(n) $\approx$ 2m(K$^\pm$) $\approx$ 2m(K$^0$). The $\Xi^0_c$ 
baryon which is a doublet of one mode has also spin 1/2. Composite particles 
which consist of a doublet of one mode plus one or two other single modes 
have spin 1/2, as the 
 $\Sigma^0$, $\Xi^0$\, and $\Lambda_c^+$, $\Sigma_c^0$, $\Omega_c^0$ 
baryons do. The only particle which seems to be the triplet of a single mode,
 the $\Omega^-$ baryon with m($\Omega^-$) $\approx$ 3m($\eta$), has
 spin 3/2. It appears that the relation between
the spin and the oscillation modes of the particles is straightforward.
 
  In the standing wave model  the $\pi^0$ and $\eta$ mesons consist 
of N = 2.85$\cdot10^9$ standing electromagnetic waves, each with its
 own frequency.
Their oscillations are longitudinal. The longitudinal oscillations of
 frequency $\nu_i$ in the $\pi^0$ and 
$\eta$   mesons do not have angular momentum or 
 $\sum_{i}j(\nu_i)$ = 0, with the index running from 0 to N. 
 Longitudinal oscillations 
 cannot cause an intrinsic angular momentum because for longitudinal 
 oscillations $\vec{r}\,\times\,\vec{p}$ = 0.

   Each of the standing waves in the $\pi^0$ and $\eta$ mesons may, 
on the other hand, have spin s = 1 of its 
own, because circularly polarized electromagnetic waves have an angular 
momentum as was first suggested by Poynting [48] and verified by, 
among others, Allen [49]. The creation of the $\pi^0$\,meson in the reaction 
$\gamma \,+\, p \rightarrow \pi^0$ + p and conservation of angular 
momentum dictates that the sum of the angular momentum 
vectors of the N electromagnetic waves in the $\pi^0$\,meson is 
zero, $\sum_ij(s_i)$ = 0.  Either the 
sum of the spin vectors of the electromagnetic 
waves in the $\pi^0$\,meson is zero, or each electromagnetic wave in the 
$\pi^0$\,meson has zero spin which would mean that they are linearly 
polarized. Linearly polarized electromagnetic waves are not expected to 
have angular momentum. That this is actually so was proven by Allen [49]. 
 Since the longitudinal  
oscillations in the $\pi^0$ and $\eta$ mesons do not have 
angular momentum and since the sum of the spin vectors $s_i$ of the 
electromagnetic  waves is zero, the intrinsic angular 
momentum of the $\pi^0$ and $\eta$ mesons is zero, or
\begin{equation} \sum_{i}\,j(\nu_i) +\sum_{i} \,j(s_i) = 0\quad 
(0\,\le\,i\le \mathrm{N})\,.\end{equation}
In the standing wave model the $\pi^0$ and $\eta$ mesons 
do not have an intrinsic angular momentum or spin, as it must be.

  We now consider particles such as the $\Lambda$ baryon 
which consist of superpositions of two circular oscillations of equal
 amplitudes and of frequencies $\omega$ and $\mathrm{-}\,\omega$,
 $|\mathrm{-}\,\omega |$ = $\omega$, at each of the N points of the 
lattice. The oscillations in the particles are coupled what we have 
marked in Tables 1,2 by the $\cdot$ sign. These particles contain 
N circular oscillations, each with its own frequency and each 
having an angular momentum of $\hbar$/2 as we will see.

  The superposition of two perpendicular linearly polarized traveling 
waves of equal amplitudes and frequencies shifted in phase by $\pi$/2 
leads to a circular wave with 
the constant angular momentum j = $\hbar$. The total 
energy of a traveling wave is the sum of the potential and the 
kinetic energy. In a traveling wave the 
kinetic energy is always equal to the potential energy. From 

\begin {equation} E_{pot} + E_{kin} = E_{tot} = \hbar\omega\,,\end 
{equation}
\noindent
follows \begin{equation} E_{tot} = 2E_{kin} = 
2\frac{\Theta\omega^2}{2}\, = \hbar\omega 
,\end{equation}
\noindent
with the moment of inertia $\Theta$. It follows that the angular momentum 
j is
 
\begin{equation} j = \Theta\omega = \hbar\,.\end{equation}
\noindent
   This applies to a traveling wave  and corresponds to spin s = 1, or
 to a circularly polarized photon.

  We now add to one monochromatic circular  
oscillation with frequency $\omega$ a  second circular oscillation 
with $\mathrm{-}\,\omega$  of the same absolute value as $\omega$ but 
 shifted in phase by $\pi$,  having the same amplitude, as we have done 
 in [45]. Negative frequencies are permitted solutions of the equations for the
 lattice oscillations. In other words we consider the circular oscillations

\begin{equation} x(t) = exp[i\omega t] + exp[-\,i(\omega t + 
\pi)]\,,\end{equation}

\begin{equation} y(t) = exp[i(\omega t + \pi/2)] + exp[-\,i(\omega t + 
                 3\pi/2)]\,\,.
\end{equation}
\noindent
This can also be written as 
\begin{equation} x(t) = exp[i\omega t] - exp[-i\omega t]\,,\end{equation}

\begin{equation} y(t) = i\cdot(exp[i\omega t] + exp[-i\omega t] )\,.\end{equation}
If we replace \emph{i} in the Eqs. above by $-$\,\emph{i} we have a 
circular oscillation turning in opposite direction.
The energy of the superposition of the two oscillations is the sum of the 
energies of both individual oscillations, and since in circular oscillations
 E$_{kin}$ = E$_{pot}$  we have according to Eq.(49) 

\begin{equation} 4E_{kin} = 4\Theta\omega^2/2 = E_{tot} = 
\hbar\omega\,,\end{equation}
\noindent
from which follows that the circular oscillation has an angular momentum

\begin{equation} j = \Theta\omega = \hbar/2\,. \end{equation}
\noindent
The superposition of two circular monochromatic  oscillations of equal
 amplitudes and frequencies $\omega$ and $\mathrm{-}\,\omega$ satisfies 
the necessary condition for spin \\s = 1/2 that the angular momentum is 
j = $\hbar$/2.

   The standing wave model treats the $\Lambda$ 
baryon, which has spin s = 1/2 and a mass m($\Lambda$) = 
1.0190\,$\cdot$\,2m($\eta$), as the superposition of two  particles of the 
same type 
with N standing electromagnetic waves. The waves are circular because 
they are the superposition of two circular waves with the same
 absolute value of the frequency and the same amplitude.  The angular 
momentum  vectors  of all circular waves in the lattice cancel,   
except for the wave at the center of the crystal. Each 
oscillation with frequency $\omega$ at $\phi$\,$>$\,0 has at its mirror 
position $\phi$\,$<$\,0 a wave with the frequency $-\,\omega$, 
which has a negative angular momentum since j = mr$^2\omega$ and
$\omega = \omega_0\phi$. 
Consequently the angular momentum vectors of both waves cancel. 
The center of the lattice oscillates, as all other lattice points do, but with 
the frequency $\nu$(0) which is determined by the longest 
possible wavelength, which is twice the sidelength d of the lattice, so 
$\nu$(0) = c/2d. As the other circular waves in the lattice the 
circular wave at the center has the angular momentum $\hbar$/2 according
 to Eq.(56). The angular momentum of the center wave is the only angular
momentum which is not canceled by an oscillation of opposite circulation.    
 
    Consequently the net angular momentum of the N
 circular oscillations in the lattice which are  superpositions 
of two oscillations reduces to the angular momentum
 of the center oscillation and is $\hbar$/2.  Since the 
circular lattice oscillations in the $\Lambda$ baryon are the 
only possible contribution to an angular momentum the 
intrinsic angular momentum of the $\Lambda$ baryon is $\hbar$/2 or
\begin{equation} j(\Lambda) = \sum_i\,j(\nu_i) = j(\nu_0) =  
\hbar/2\,.\end{equation}             
 We have thus explained that the $\Lambda$ and likewise
 the $\Xi_c^0$ baryon satisfy the necessary condition 
that j = $\hbar$/2 for s = 1/2. The intrinsic angular momentum of the 
$\Lambda$ baryon is the consequence of the superposition of two 
circular oscillations of the same amplitude and the same absolute value
of the frequency.

   The other particles of the $\gamma$-branch, the 
$\Sigma^0$, $\Xi^0$, $\Lambda^+_c$, $\Sigma^0_c$ and $\Omega^0_c$
 baryons are composites of a baryon with spin 1/2 plus one or two $\pi$\,mesons
 which do not have spin. Consequently the spin of these particles is 
1/2. The spin of all particles of the $\gamma$-branch, exempting the 
spin of the $\Omega^-$\,baryon, has thus been explained.
 For an explanation 
of  s($\Sigma^{\pm,0}$)  = 1/2 and of s($\Xi^{-,0}$)  = 1/2, regardless
 whether the particles  are charged or neutral, we refer to [45].    

\section{The spin of the particles of the $\nu$-branch}

The characteristic particles of the neutrino-branch are the 
$\pi^\pm$\,mesons 
which have zero spin. At first glance it seems to be odd that the 
$\pi^\pm$\,mesons do not have spin, because it seems that the 
$\pi^\pm$\,mesons should have spin 1/2 from the spin of the charges 
e$^\pm$ in $\pi^\pm$. However that is not the case. The 
solution of this puzzle is in the composition of the $\pi^\pm$\,mesons 
which are, according to the standing wave model, made of a lattice of 
neutrinos and antineutrinos (Fig.\,2) each having spin 1/2, the lattice oscillations,
 and an electrical charge.

  The longitudinal oscillations in the neutrino lattice of the $\pi^\pm$\,mesons 
do not cause an angular momentum, $\sum_i\,j(\nu_i)$ = 0, as it was with the $\pi^0$\,meson. In the cubic lattice of N = O($10^9$) neutrinos and 
antineutrinos of the $\pi^\pm$\,me-sons the spin of 
 nearly all neutrinos and antineutrinos must cancel because conservation 
of angular momentum during the creation of the particle requires that the 
total angular momentum around a central axis is $\hbar$/2. In 
fact the spin vectors of all but the neutrino or antineutrino in the 
center of the lattice cancel. In order for this to be so
 the spin vector of any particular neutrino in the lattice 
has to be opposite to spin vector of the neutrino at its 
mirror position. As is well-known only left-handed neutrinos and 
right-handed antineutrinos exist. From $\nu$ = $\nu_0\phi$ (Eq.13) follows 
that the direction of motion of the neutrinos in e.g.\,\,the upper right 
quadrant ($\phi$\,$>$\,0) is opposite to the direction of motion in the 
lower left quadrant ($\phi$\,$<$\,0). Consequently the spin vectors of all 
neutrinos or antineutrinos in opposite quadrants are opposite and cancel.
 The only angular 
momentum remaining from the spin of the neutrinos of the lattice is the 
angular momentum of the neutrino or antineutrino at the center of the 
lattice which does not have a mirror particle. 
Consequently the electrically neutral neutrino lattice consisting of 
N$^\prime$/2 neutrinos and N$^\prime$/2 antineutrinos and the center
 particle, each with spin s($n_i$) = 1/2, has an intrinsic 
angular momentum j = $\sum_i\,j(n_i)$ = \emph{j(n$_0$}) = $\hbar$/2.

   But electrons or positrons added to the neutral 
neutrino lattice have spin 1/2. If the spin of the electron or positron 
added to the neutrino lattice is opposite to the spin of the neutrino or
 antineutrino in the center of the lattice then the net spin of the
 $\pi^+$ or $\pi^-$ mesons is zero, or
 \begin{equation} j(\pi^\pm) = \sum_i\,j(n_i) + j(e^\pm) = j(n_0) + j(e^\pm) = 0
\,\quad(0\leq i \leq \mathrm{N})\,.
\end{equation}
It is important for the understanding of the structure of 
the $\pi^\pm$\,mesons to realize that s($\pi^\pm$) = 0 can only be 
explained if the $\pi^\pm$\,mesons consist of a \emph{neutrino lattice} 
to which an electron or positron is added whose spin is opposite to the 
net 
spin of the neutrino lattice. \emph{Spin 1/2 of the electric charges can only be 
canceled by something that has also spin 1/2, and the only
 conventional choice for that is a single neutrino}.

   \emph{The spin, the mass and the decay of $\pi^\pm$ require that the 
$\pi^\pm$\,mesons are made of a neutrino lattice and an electrical charge}.

   The spin of the K$^{\pm}$\,mesons is zero. With the spin of the
 K$^\pm$\,mesons we 
encounter the same oddity we have just observed with the spin of the 
$\pi^\pm$\,mesons, namely we have a particle which carries an electrical 
charge with spin 
1/2, and nevertheless the particle does not have spin. The explanation 
of s(K$^\pm$) = 0 follows the same lines as the explanation of 
the spin of the $\pi^\pm$\,mesons. In the standing wave model the
 K$^\pm$\,mesons are described by the state (2.)$\pi^\pm$ + $\pi^0$,
 that means by the second 
mode of the $\pi^\pm$\,mesons plus a $\pi^0$\,meson.
 The second mode of the longitudinal oscillations of a neutral neutrino 
lattice  
 does not have a net intrinsic angular momentum $\sum_i\,j(\nu_i)$ = 0. But
 the spin of the neutrinos contributes an angular momentum $\hbar$/2, 
which 
originates from the neutrino or antineutrino in the center of the lattice, 
just as it is with the neutrino lattice in 
the $\pi^\pm$\,mesons, so $\sum_i\,{ j(n_i)} = \hbar/2$. Adding 
an electric charge with a spin opposite to the net intrinsic angular 
momentum of the neutrino lattice oscillations creates the charged 
(2.)$\pi^\pm$ mode which has zero spin

\begin{equation}j((2.)\pi^\pm) = \sum_i\,{j(n_i)} + j(e^\pm) = j(n_0) + j(e^\pm) = 0\,.
\end{equation}
 As discussed in Section 6 it is 
necessary to add a $\pi^0$\,meson to the second mode of the 
$\pi^\pm$\,meson in 
order to obtain the correct mass and the correct decays of the 
K$^\pm$\,mesons. Since the $\pi^0$\,meson 
does not have spin the addition of the $\pi^0$\,meson does not add to the 
intrinsic angular momentum of the K$^\pm$\,mesons. So 
 s(K$^\pm$) = 0 as it must be.

  The explanation of s = 0 of the K$^0$ and 
$\overline{{\mathrm{K}}^0}$\,mesons described by 
the state (2.)$\pi^\pm$ + $\pi^\mp$  is different. The 
 oscillations of the second mode of $\pi^\pm$ as 
well as of the basic $\pi^\mp$ mode do not create an angular momentum, 
$\sum_i\,j(\nu_i)$ = 0. The second mode of the $\pi^\pm$\,mesons,
or the (2.)$\pi^\pm$ state, and the 
basic $\pi^\mp$ mode each have N$^\prime$/2 neutrinos and 
N$^\prime$/2 antineutrinos plus a center neutrino or antineutrino, so the 
number of all neutrinos and antineutrinos in the sum of both states, 
the K$^0$\,meson, is 2N. Since the size of the lattice of the 
K$^\pm$\,mesons and 
the K$^0$\,mesons is the same it follows that two neutrinos are at each 
lattice point of the K$^0$ or $\overline{{\mathrm{K}}^0}$\,mesons. We 
assume that Pauli's exclusion principle applies 
for neutrinos as well. Consequently each neutrino at each lattice point must 
share its location with an antineutrino. That means that the contribution 
of the spin of all neutrinos and antineutrinos to the intrinsic angular momentum
 of the K$^0$\,meson is zero or $\sum_i\,j(2n_i)$ = 0.  The 
sum of the spin vectors of the two opposite charges in the K$^0$ and
$\overline{{\mathrm{K}}^0}$\,mesons, 
or in the (2.)$\pi^\pm$ + $\pi^\mp$ state, is also 
zero. Since neither the lattice oscillations nor the spin of the 
neutrinos and antineutrinos 
nor the  electric charges contribute an angular momentum 
\begin{equation} j(\mathrm{K}^0) = \sum_i\,j(2n_i) + j(e^+ + e^-) = 0\,.
\end{equation}
 The intrinsic angular 
momentum of the K$^0$ and $\overline{{\mathrm{K}}^0}$\,mesons
 is zero, or s(K$^0$,$\overline{{\mathrm{K}}^0}$) = 0, as it 
must be. In simple terms, since the structure of K$^0$ is 
(2.)$\pi^+$ + $\pi^-$, the spin of K$^0$ is the sum of the spin of  
(2.)$\pi^+$ and of $\pi^-$, both of which do not have spin. 
It does not seem possible to arrive at
 s(K$^0$,$\overline{{\mathrm{K}}^0}$) = 0  if both particles 
do not contain the N pairs of neutrinos and antineutrinos required by the
 (2.)$\pi^\pm$ + $\pi^\mp$ state which we have suggested in Section 6.  

   In the case of the neutron 
one must wonder how it comes about that a particle which seems to be the 
superposition of two particles without spin ends up with spin 1/2. The 
neutron,
 which has  a mass $\approx$ 2m(K$^\pm$) or 2m(K$^0$), is either the 
superposition of a K$^+$ and a K$^-$ meson or of a K$^0$ and a 
$\overline{\mathrm{K}^0}$ meson. The intrinsic angular 
momentum of the superposition of K$^+$ and K$^-$ is either 0 or $\hbar$, 
which means that the neutron cannot be the superposition of K$^+$ 
and K$^-$. For a proof of this statement we must refer to [45].

   On the other hand the neutron seems to be the superposition of a 
K$^0$  and a $\overline{\mathrm{K}^0}$\,meson. A significant change 
in the lattice occurs when a K$^0$ and a  $\overline{\mathrm{K}^0}$ meson
 are superposed. Since each K$^0$\,meson contains N neutrinos 
and N antineutrinos, as we 
discussed before in context with the spin of K$^0$, the number of all 
neutrinos and antineutrinos in superposed K$^0$  and 
$\overline{\mathrm{K}^0}$ lattices is 4N. Since the size of the 
lattice of the proton as well of the neutron is the same as the size of 
K$^0$ each of the N lattice 
points of the neutron now contains four neutrinos, a muon neutrino and
 an anti-muon neutrino as well as an 
electron neutrino and an anti-electron neutrino. The 
$\nu_\mu,\bar{\nu}_\mu,\nu_e,\bar{\nu}_e$ quadrupoles  
oscillate just like individual neutrinos do because we learned from Eq.(7)
that the ratios of the oscillation frequencies are independent 
of the mass as well as of the interaction constant between the lattice 
points. In the neutrino quadrupoles the spin of the 
neutrinos and antineutrinos cancels, $\sum_i\,j(4n_i)$ = 0. The
 superposition of two circular neutrino lattice oscillations,  that means the 
circular oscillations of frequency $\nu_i$,  contribute as before the 
angular momentum of the center circular oscillation, so 
$\sum_i\,j(\nu_i) = \hbar/2$. 
The spin and charge of the four electrical charges hidden in the sum of the 
K$^0$ and $\overline{\mathrm{K}^0}$ mesons cancel.  
It follows that the intrinsic angular momentum of a neutron created by the 
superposition of  a K$^0$  and a $\overline{\mathrm{K}^0}$ meson comes
 from the circular neutrino lattice oscillations only and is 
\begin{equation} j(\mathrm{n}) = \sum_i\,j(\nu_i) + \sum_i\,j(4n_i) + j(4e^\pm) = 
 \sum_i\,j(\nu_i) = j(\nu_0) = \hbar/2\,,
\end{equation}
as it must be. In simple terms, the spin of the neutron originates from 
the superposition of two circular neutrino lattice 
oscillations  with the frequencies $\omega$ and $\mathrm{-}\,\omega$
 shifted in phase by $\pi$, which 
produces the angular momentum $\hbar$/2.

   The spin of the proton is 1/2 and is unambiguosly defined by the decay 
of the neutron n $\rightarrow$ p + e$^-$ + $\bar{\nu}_e$. We have suggested
 in Section 7 that 3/4$\cdot$N$^\prime$ anti-electron neutrinos of the neutrino
 lattice of the neutron
 are removed in the $\beta$-decay of the neutron and that N$^\prime$/4
 anti-electron neutrinos leave with the emitted electron. The intrinsic 
angular momentum of the proton originates then from the spin of the 
central $\nu_\mu\bar{\nu}_\mu\nu_e$ triplet, from the spin of
 the e$^+$e$^-$e$^+$ triplet which is part of the remains of the neutron,
 and from the angular momentum of the center of the lattice oscillations 
with the superposition of two circular oscillations. The spin of the 
central $\nu_\mu\bar{\nu}_\mu\nu_e$ triplet
is canceled by the spin of the e$^+$e$^-$e$^+$ triplet. According to the
 standing wave model the intrinsic angular momentum of the proton is

\begin{equation}
 j(\mathrm{p}) = j(\nu_\mu\bar{\nu}_\mu\nu_e)_0\,  +  j(e^+e^-e^+)
+ j(\nu_0) = j(\nu_0) = \hbar/2\,, 
\end{equation}
\noindent as it must be.
                         
   The other mesons of the neutrino branch, the D$^{\pm,0}$ and 
D$_s^\pm$ mesons, both having zero spin, 
are superpositions of a proton and an antineutron of opposite spin, or of 
their antiparticles, or of a neutron and an antineutron of opposite spin in 
D$^0$.  The spin of D$^\pm$ and D$^0$ does therefore not pose a new 
problem. The spin of D$^\pm_s$  is explained in [39].

   For an explanation of the spin of $\mu^\pm$ we refer to [50]. Since all muon  
or anti-muon neutrinos have been removed from the $\pi^\pm$ lattice in the
$\pi^\pm$ decay it follows that a neutrino vacancy is at the center of the
 $\mu^\pm$ lattice. Without a neutrino
 in the center of the lattice the sum of the spin vectors of all neutrinos in the 
$\mu^\pm$ lattice is zero. However the $\mu^\pm$\,mesons consist 
of the neutrino lattice plus an electric charge whose spin is 1/2. The spin of the 
$\mu^\pm$\,mesons originates from the spin of the electric charge carried
by the $\mu^\pm$\,mesons and is consequently s($\mu^\pm$) = 1/2. The 
same considerations apply for the spin of $\tau^\pm$, s($\tau^\pm$) = 1/2.   

   An explanation of the spin of the mesons and baryons can only be valid 
if the same explanation also applies to the antiparticles of these 
particles whose spin is the $\emph{same}$ as that of the ordinary particles. 
 The antiparticles of the $\gamma$-branch consist of electromagnetic waves 
whose frequencies differ from the frequencies of the ordinary particles only by
their sign. The angular momentum of the superposition of two circular oscillations 
with $\mathrm{-}\,\omega$ and $\omega$  has the same angular momentum 
as the superposition of two circular oscillations 
 with frequencies of opposite  sign, as in $\Lambda$. Consequently the
 spin of the antiparticles of the $\gamma$-branch is the same as the spin of the 
ordinary particles of the $\gamma$-branch. The same considerations apply to 
the circular neutrino lattice oscillations which cause the spin of the neutron, the only 
particle of the $\nu$-branch which has spin. In the standing wave model the spin 
of the neutron and the antineutron are the same.

   From the foregoing we arrive at an understanding of the reason for the astonishing
 fact that the intrinsic angular momentum or spin of the particles is independent of
the mass of the particles, as exemplified by the spin $\hbar$/2 of the electron
being the same as the spin $\hbar/2$ of the proton, notwithstanding the fact that
 the mass of the proton is 1836 times larger than the mass of the electron.  
However, in our model, the spin of the particles including the electron 
is determined solely by the angular
 momentum $\hbar$/2 of the center point of the lattice, the other angular momentum 
vectors in the particles cancel. Hence the mass of the particles contained in the other 
$10^9$  lattice points is inconsequential for the intrinsic angular momentum of the 
particles. That does indeed mean that the spin of the particles is independent of 
the mass of the particles.        

\section*{Conclusions}

   We conclude that the standing wave model solves a number of problems 
for which an answer heretofore has been hard to come by. Only photons, 
neutrinos, charge and the weak nuclear force are needed to explain the
  masses of the stable mesons and baryons and of the leptons. We can also 
explain the spin of the baryons and the absence of spin in the mesons,
 and the spin of the $\mu^\pm$ and $\tau^\pm$\,mesons as well.  

\vspace{1cm}

   {\bfseries Acknowledgments}. I gratefully acknowledge the contributions
 
of 
Dr. T. Koschmieder to this study. I thank Professor A. Gsponer \\ 
\indent for information about the history of the integer 
multiple rule.

\section*{References}

\noindent
[1] Gell-Mann,M. Phys.Lett.B {\bfseries111},1 (1964).

\smallskip
\noindent
[2] Hagiwara,K. et al. Phys.Rev.D {\bfseries 66},010001 (2002).

\smallskip
\noindent
[3] Witten,W. Physics Today {\bfseries 46},\#4,24 (1996). 

\smallskip
\noindent
[4] El Naschie,M.S. Chaos,Sol.Frac. {\bfseries 14},649 (2002).

\smallskip
\noindent
[5] El Naschie,M.S. Chaos,Sol.Frac. {\bfseries 14},369 (2002).

\smallskip
\noindent
[6] El Naschie,M.S. Chaos,Sol.Frac. {\bfseries 16},353 (2003).

\smallskip
\noindent
[7] El Naschie,M.S. Chaos,Sol.Frac. {\bfseries 16},637 (2003).

\smallskip
\noindent
[8] Feyman,R.P. \emph{QED. The Strange Theory of Light and Matter}.\\
\indent\,\, Princeton University Press, p.152 (1985).

\smallskip
\noindent
[9] Koschmieder,E.L. Chaos,Sol.Frac. {\bfseries18},1129  (2003).
 
\smallskip
\noindent
[10] Nambu,Y. Prog.Th.Phys. {\bfseries 7},595 (1952).

\smallskip
\noindent
[11] Fr\"ohlich,H. Nucl.Phys. {\bfseries 7},148 (1958). 

\smallskip
\noindent
[12] Koschmieder,E.L. and Koschmieder,T.H.\\
\indent Bull.Acad.Roy.Belgique {\bfseries X},289 (1999),\\
\indent http://arXiv.org/hep-lat/0002016 (2000).

\smallskip 
\noindent
[13] Wilson,K.G. Phys.Rev.D {\bfseries10},2445 (1974).

\smallskip
\noindent
[14] Born,M. and v.\,Karman,Th. Phys.Z. {\bfseries13},297 (1912).

\smallskip
\noindent
[15] Blackman,M. Proc.Roy.Soc.A {\bfseries148},365;384 (1935).

\smallskip
\noindent
[16] Blackman,M. Handbuch der Physik VII/1, Sec.12 (1955).

\smallskip
\noindent
[17] Born,M. and Huang,K. \emph{Dynamical Theory of Crystal Lattices},\\ 
\indent \,\,(Oxford) (1954).

\smallskip
\noindent
[18] Maradudin,A. et al. \emph{Theory of Lattice Dynamics in the 
Harmonic\\ 
\indent \,\, Approximation}, Academic Press, 2nd edition, (1971).

\smallskip
\noindent
[19] Ghatak,A.K. and Khotari,L.S. \emph{An introduction to Lattice 
Dynamics},\\
\indent \,\, Addison-Wesley, (1972).

\smallskip
\noindent
[20] Rekalo,M.P., Tomasi-Gustafson,E. and Arvieux,J. Ann. Phys.\\
 \indent\,\, {\bfseries295},1 (2002).

\smallskip
\noindent
[21] Schwinger,J. Phys.Rev. {\bfseries128},2425 (1962).

\smallskip
\noindent
[22] Sidharth,B.J. \emph{The Chaotic Universe}, p.49\\
\indent\,\, Nova Science Publishers, New York						 (2001).

\smallskip
\noindent
[23] Perkins,D.H. \emph{Introduction to High-Energy Physics},\\
\indent \,\,Addison Wesley, (1982).

\smallskip
\noindent
[24] Rosenfelder,R. Phys.Lett.B {\bfseries479},381 (2000).

\smallskip
\noindent
[25] Born,M. Proc.Camb.Phil.Soc. {\bfseries36},160 (1940).

\smallskip
\noindent
[26] Koschmieder,E.L. http://arXiv.org/hep-lat/0005027 (2000).

\smallskip
\noindent
[27] Melnikov,K. and van Ritbergen,T. Phys.Rev.Lett. {\bfseries84},1673 (2000).

\smallskip
\noindent
[28] Liesenfeld,A. et al. Phys.Lett.B {\bfseries468},20 (1999).

\smallskip
\noindent
[29] Bernard,V., Kaiser,N. and Meissner,U-G.\\
\indent  \,\,Phys.Rev.C {\bfseries62},028021 (2000).

\smallskip
\noindent
[30] Eschrich,I. et al. Phys.Lett.B {\bfseries522},233 (2001).

\smallskip
\noindent
[31] Sommerfeld,A. \emph{Vorlesungen \"{u}ber Theoretische Physik},\\
\indent  \,\,Bd.V, p.56 (1952).

\smallskip
\noindent
[32] Debye,P. Ann.d.Phys. {\bfseries39},789 (1912).

\smallskip
\noindent
[33] Bose,S. Z.f.Phys. {\bfseries26},178 (1924).

\smallskip
\noindent
[34] Bethe,H. Phys.Rev.Lett. {\bfseries58},2722 (1986).

\smallskip
\noindent
[35] Bahcall,J.N. Rev.Mod.Phys. {\bfseries59},505 (1987).

\smallskip
\noindent
[36] Fukuda,Y. et al. Phys.Rev.Lett. {\bfseries81},1562 (1998), {\bfseries90},171302 (2003).

\smallskip
\noindent
[37] Ahmad,Q.R. et al. Phys.Rev.Lett. {\bfseries87},071301 (2001).

\smallskip
\noindent
[38] Lai,A. et al. Eur.Phys.J. C {\bfseries30},33 (2003).

\smallskip
\noindent
[39] Koschmieder,E.L. http://arXiv.org/physics/0309025 (2003).

\smallskip
\noindent
[40] Barut,A.O. Phys.Rev.Lett. {\bfseries 42},1251 (1979).

\smallskip
\noindent
[41] Gsponer,A. and Hurni,J-P. Hadr.J. {\bfseries 19},367 (1996).

\smallskip
\noindent
[42] Born,M. and Stern,O. Sitzungsber.Preuss.Akad.Wiss.\\ 
\indent  \,\,{\bfseries33},901 (1919).

\smallskip
\noindent
[43] Barut,A.O. Phys.Lett.B {\bfseries73},310 (1978).

\smallskip
\noindent
[44] Uhlenbeck,G.E. and Goudsmit,S. Naturwiss. {\bfseries13},953 (1925). 

\smallskip
\noindent
[45] Koschmieder,E.L. Hadr.J. {\bfseries26},471 (2003),\\
\indent\,\,http://arXiv.org/physics/0301060 (2003).

\smallskip
\noindent
[46] Jaffe,R.L. Phil.Trans.Roy.Soc.Lond.A {\bfseries359},391 (2001).

\smallskip
\noindent
[47] G\"ockeler,M. et al. Phys.Lett.B {\bfseries545},112 (2002).

\smallskip
\noindent
[48] Poynting,J.H. Proc.Roy.Soc.A {\bfseries82},560 (1909).

\smallskip
\noindent
[49] Allen,J.P. Am.J.Phys. {\bfseries34},1185 (1966).

\smallskip
\noindent
[50] Koschmieder,E.L. http://arXiv.org/physics/0308069 (2003).
\bigskip
\bigskip

\newpage

{\bfseries The magnetic moment of the neutron}
\bigskip

   In the same way we can explain the magnetic moment of the neutron which 
has a magnetic moment with a g-factor g(n) = - 3.826 = - 0.956$\cdot$4. This
magnetic moment must originate from a pair of electric charges since the 
g-factor of a single electric charge is $\cong$ 1.00116$\cdot$2. At least two
 electric charges of opposite sign must be within the neutron to have any
 magnetic moment at all. However, in the standing wave model four charges 
are in the neutron, two each of opposite sign. In order to have a magnetic
 moment twice the moment of either e$^+$ or e$^-$ the spin of two charges
 must be parallel, whereas the spin of the two other charges must be antiparallel
so that their magnetic moments cancel, or there must be two opposite charges
 with opposite spin. As discussed in the preceding paragraph
 the magnetic moment of the proton is caused by the parallel spin of two e$^+$
and the opposite spin of one e$^-$, causing a moment three times the 
moment of one charge. These three charges carry N/4 electron neutrinos each,
because all anti-electron neutrinos have been removed from the neutron lattice 
in its decay, as mentioned before. As follows from the decay n $\rightarrow$
 p + e$^-$ + $\bar{\nu}_e$ there must be an additional e$^-$ in the neutron
 as compared to the proton. The additional e$^-$ carries N/4 anti-electron
 neutrinos because of conservation of angular momentum. Consequently this
 electron has a spin opposite to the spin of the e$^-$ in the proton. The 
magnetic moments of the two electrons with opposite spin cancel. However 
two positive charges e$^+$ having the same spin are also in the neutron, as
 they are in the proton. They create the magnetic moment of the neutron and
 have a g-factor $\cong$ 4.

   There seems to be an excess angular momentum caused by the parallel spin of
 the two e$^+$. However this is taken care of by the spin of the anti-electron neutrino 
which comes with the neutron decay and by the spin of the N/4 electron neutrinos
 not bound in the electric charges in the neutron. Summing up, the magnetic
 moment of the neutron is caused by the parallel spin of a pair of the four 
electrical charges which are in the neutron according to our model. The magnetic
 moment of the other pair of charges cancel because their spin is antiparallel.

\end{document}